\documentclass[10pt,twocolumn,twoside]{IEEEtran}
\usepackage{amsbsy,amsmath,amsfonts,amssymb,amsbsy,subfigure,verbatim,color,array}
\usepackage{bm,cite,graphicx,psfrag,pstricks,theorem,tikz,times,url}
\usepackage{wrapfig}
\usetikzlibrary{shapes,snakes,calendar,matrix,backgrounds,folding}
\usepackage[english]{babel}
\usetikzlibrary{arrows,automata}
\usetikzlibrary{calc} 
\usetikzlibrary{decorations.pathmorphing} 

\newcolumntype{M}{>{\centering\arraybackslash}m{\dimexpr.45\linewidth-2\tabcolsep}}

\newcommand{\bi}{\begin{itemize}}
\newcommand{\ei}{\end{itemize}}
\newcommand{\ben}{\begin{enumerate}}
\newcommand{\een}{\end{enumerate}}

\newcommand{\bc}{\begin{cases}}
\newcommand{\ec}{\end{cases}}
\newcommand{\bd}{\begin{description}}
\newcommand{\ed}{\end{description}}

\newcommand{\be}{\begin{equation}}
\newcommand{\ee}{\end{equation}}
\newcommand{\bea}{\begin{eqnarray}}
\newcommand{\eea}{\end{eqnarray}}

\newcommand{\meas}[1]{\,\,\,\mathrm{#1}}

\newtheorem{definition}{Definition}

\theoremstyle{plain}
\newtheorem{remark}{Remark}

\newcommand{\nn}{\nonumber}

\graphicspath{%
     {./Figures/}
}

\begin{document}

\title{A Stochastic Model for Electron Transfer\\in Bacterial Cables}
\author{
Nicol\`{o}~Michelusi,~Sahand~Pirbadian,
Mohamed~Y.~El-Naggar~and~Urbashi~Mitra
\thanks{N. Michelusi and U. Mitra are with the Ming Hsieh Department of Electrical Engineering, University of Southern California, Los Angeles, USA;
M. Y. El-Naggar and S. Pirbadian are with the Department of Physics and Astronomy, University of Southern California, Los Angeles, USA; emails:
\tt{\{michelus,spirbadi,mnaggar,ubli\}@usc.edu}
}
\thanks{N. Michelusi and U. Mitra acknowledge support from one or all of these grants:  
ONR N00014-09-1-0700, CCF-0917343, CCF-1117896, CNS-1213128, AFOSR FA9550-12-1-0215, and DOT CA-26-7084-00.
S. Pirbadian and M. Y. El-Naggar acknowledge support from NASA Cooperative Agreement NNA13AA92A 
and grant DE-FG02-13ER16415 from the Division of Chemical Sciences, Geosciences, and Biosciences, Office of Basic Energy Sciences of the US Department of Energy.
N. Michelusi is in part supported by AEIT (Italian association of electrical engineering) through the research scholarship "Isabella Sassi Bonadonna 2013".
}
\thanks{Parts of this work have appeared in \cite{CISS}.}
\vspace{-5mm}
}
\maketitle
\begin{abstract}
Biological systems are known to communicate by diffusing chemical signals in the surrounding medium. 
However, most of the recent literature has neglected the \emph{electron transfer} mechanism occurring amongst living cells, and its role in cell-cell communication.
Each cell relies on a continuous flow of electrons from its electron donor to its electron acceptor through the  electron transport chain to produce energy in the form of the molecule adenosine triphosphate, and to sustain the cell's vital operations and functions.
While the importance of biological electron transfer is well-known for individual cells, the past decade has also brought about remarkable discoveries of multi-cellular microbial communities that transfer electrons between cells and across centimeter length scales, \emph{e.g.}, biofilms and multi-cellular bacterial cables. 
These experimental observations open up new frontiers in the design of electron-based communications networks in microbial communities,
 which may coexist with the more well-known communication strategies based on molecular diffusion, while benefiting from a much shorter communication delay.
This paper develops a stochastic model that links the electron transfer mechanism to the energetic state of the cell. The model is also extensible to larger communities, by allowing for electron exchange between neighboring cells.
  Moreover, the parameters of the stochastic model are fit to experimental data available in the literature, and are shown to provide a good fit.
\end{abstract}
%
\section{Introduction}
Biological systems are known to communicate by diffusing chemical signals in the surrounding medium. One example is \emph{quorum sensing} \cite{Bassler,Visick15082005,Nealson}, where the concentration of certain signature chemical compounds emitted by the bacteria is used to estimate the bacterial population size, so as to simultaneously activate a certain collective behavior.
{More recently, molecular communication has been proposed as a viable communication scheme for nanodevices and nanonetworks,
and is under IEEE standards consideration \cite{P1906}.}
The performance evaluation, optimization and design of molecular communications systems opens up new challenges in the information theory \cite{bush2010nanoscale,Nakano,Akyildiz1,Rose}.
The achievable capacity of the chemical channel using molecular communication
is investigated in \cite{Eckford07,Kadloor}, under Brownian motion, and in \cite{Einolghozati}, under a diffusion channel.
 In \cite{Fekri}, a new 
architecture for networks of bacteria to form a data collecting network is described, and aspects such as reliability and speed of convergence of consensus are investigated.
In \cite{Arjmandi,Mosayebi}, a new molecular modulation scheme for nanonetworks is proposed and analyzed, based on the
idea of time-sharing between different types of molecules in order to effectively suppress the interference.
In \cite{Oiwa}, an in-vitro molecular communication system  is designed and,
in \cite{Kuran201086}, an energy model is proposed, based on molecular diffusion.

While communication via chemical signals has been the focus of most prior investigations,
experimental evidence on the microbial emission and response to three physical signals, \emph{i.e.}, sound waves, electromagnetic radiation and electric currents,
suggests that physical modes of microbial communication could be widespread in nature \cite{Reguera2}.
In particular, communication exploiting electron transfer in a bacterial network has previously been observed in nature \cite{Pfeffer} and in bacterial colonies in lab
\cite{Kato}. 
This multi-cellular communication is usually triggered by extreme environmental conditions, \emph{e.g.}, lack of
electron donor (ED) or electron acceptor (EA), in turn resulting in various gene expression levels and functions in different cells within the community, and enables the entire community to survive under harsh conditions.
Electron transfer is fundamental to cellular respiration: each cell relies on a continuous flow of electrons from an ED to an EA through the cell's electron transport chain (ETC) to produce energy in the form of the molecule adenosine triphosphate (ATP), and to sustain its vital operations and functions.
This strategy, known as \emph{oxidative phosphorylation}, is employed by all respiratory microorganisms.
In this regard, we can view the flow of one electron from the ED to the EA as an energy unit which is harvested from the surrounding medium to power the operations of the cell,
and stored in an internal "rechargeable battery" (energy queue,
\emph{e.g.}, see the literature on \emph{energy harvesting} for wireless communications and references therein \cite{Gunduz,MichelusiEH,Liu}).

\begin{figure}[t]
\centering
\vspace{5mm}
\includegraphics[width=.85\linewidth,trim = 5mm 3mm 15mm 9mm,clip=false]{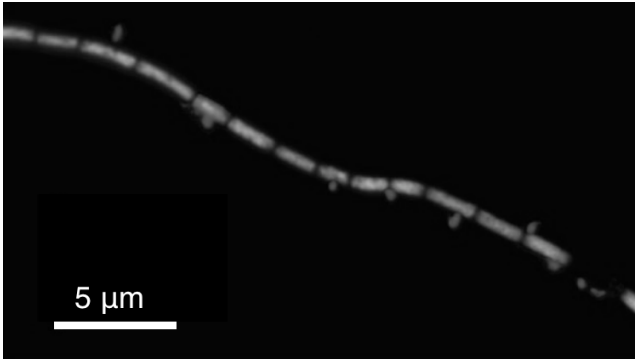}
\caption{Fluorescent image of filamentous \emph{Desulfobulbaceae}. Bacterial cells are aligned to form the cable, which couples Oxygen reduction at the marine sediment surface to sulphide reduction in deeper anoxic layers by transferring electrons along its length. From \cite{Pfeffer}.\vspace{-4mm}}
\label{figb}
\end{figure}

While the importance of biological electron transfer and oxidative phosphorylation is well-known for individual cells, the past decade has also brought about remarkable discoveries of multi-cellular microbial communities that transfer electrons between cells and across much larger length scales than previously thought \cite{Naggar}. Within the span of only a few years, observations of microbial electron transfer have jumped from nanometer to centimeter length scales, and the structural basis of this remarkably long-range transfer has evolved from recently discovered molecular assemblies known as \emph{bacterial nanowires} \cite{Naggar,Naggar2,Pirbadian}, to entire macroscopic architectures, including biofilms and multi-cellular bacterial cables, 
  consisting of thousands of cells lined up end-to-end in marine sediments \cite{Reguera,Pfeffer} (see Fig. \ref{figb}). Therein, the cells 
  in the deeper regions  of the sediment where the ED is located
  extract more electrons, while the cells in the upper layers, where Oxygen (an EA) is more abundant,
have a heightened
 transfer of electrons to the EA.
  The survival of the whole system relies on this division of labor, with the intermediate cells operating as "relays" of electrons to coordinate this collective response to the spatial separation of ED and EA.
It is worth noticing that other biological cable-like mechanisms exist in nature, enabling cell-cell communication:
 \emph{tunneling nanotubes} connect two animal cells for transport of organelles and membrane vesicles and create complex networks of interconnected cells
 \cite{Rustom};
 in the bacterial world, \emph{Myxococcus xanthus} cells form membrane tubes that connect cells to one another in order to transfer outer membrane content \cite{Remis}.

  These experimental observations raise the possibility of an electron-based communications network in microbial communities, which may coexist with the more well-known communication strategies based on molecular diffusion \cite{Naggar,Reguera2,Arjmandi}.
     For microbes, the advantage of electron-based communications is clear: in contrast to the relatively slow diffusion of whole molecules via Brownian motion, electron transfer is a rapid process that enables cells to quickly sense and respond to their environment. 
 As an example of a communications architecture based on electron transfer, consider
a system composed of an ED terminal (transmitter, or electron source) which operates as the signal encoder, an EA terminal (receiver, or electron sink)
   and the network of bacteria;
    the electron signal, encoded by the ED terminal and input into the network, is then relayed in 
  a multi-hop fashion, following
  the natural laws of electron transfer within each cell and across neighboring cells,  which this paper aims at modeling;
  the flow of electrons is finally collected at the EA terminal.
  Such an electron signal, coupled with the energetic state of each cell, can be
  "decoded" by the individual cells to activate a certain desired gene expression.
For instance, in a biofilm formed on a surface, bacteria interact with each other and with a solid phase terminal EA via electron transfer, 
  which serves \emph{both}  as a respiratory advantage and a communications scheme for bacteria to adapt to their environment.
  Additionally, electron transfer can be employed in place of molecular diffusion for quickly transporting information in nanonetworks.
In particular,  information
can be encoded in the concentration of electrons released by the encoder into the bacterial cable,
using a technique termed \emph{concentration shift keying} \cite{Kuran5962989,Arjmandi}.
   The additional challenge with respect to molecular diffusion is that
    the electron is \emph{both} an energy carrier involved in the energy production for the cell to sustain its functionalities, and an information carrier, which
    enables the transport of information between nanodevices, thus introducing additional constraints in the encoded signal.

Electron-based communication presents significant advantages, as discussed above, but this phenomenon also raises new intriguing questions. While a single cell can extract enough free energy to power life's reactions by exploiting the redox potential difference between ED oxidation and EA reduction, how can the same potential difference be used to power an entire multicellular assembly such as the \emph{Desulfobulbaceae} bacterial cables \cite{Pfeffer}? Specifically, can intermediate cells survive without access to chemical ED or EA, by exploiting the potential difference between cells in the deeper sediment (sulfide oxidizers) and cells in the oxic zone (oxygen reducers)? For a cable consisting of thousands of cells this appears unlikely, since the free energy available for an intermediate cell is inversely proportional to the total number of cells. Are additional, yet unknown, electron sources and sinks necessary to maintain the whole community? These questions necessitate flexible models that analyze emerging experimental data in order to elucidate the energetics of individual cells, as presented here, and, eventually, whole bacterial cables or biofilms.

 In order to enable the modeling and control of such microbial communications network and guide future experiments, in this paper we set out to develop a stochastic queuing theoretic model
that links electron transfer to the energetic state of the cell (\emph{e.g.}, ATP concentration or energy charge potential). We show how the proposed model can be extended to larger communities (\emph{e.g.}, cables, biofilms), by allowing for electron transfer  between neighboring cells.
 In particular,
we analyze the stochastic model for an isolated cell, which is the building block of multi-cellular networks,
 and provide an example of the application of the proposed framework to the computation of the cell's lifetime.
Finally, we design a parameter estimation framework  and fit the parameters of the model to experimental data available in the literature.
The prediction curves are compared to experimental ones, showing a good fit.
This paper represents a preliminary essential modeling step towards the design and analysis of bacterial communications networks,
and provides the ground to
 model and control bacterial interactions (\emph{e.g.}, gene expressions) induced by 
the electron transfer signal, 
and to analyze information theoretic aspects, such as the interplay between information capacity and lifetime of the cells,
as well as communication reliability  and delay.
 
This paper is organized as follows. In Sec. \ref{stochmodel}, we present a stochastic model for the cell, and 
for the interconnection of cells via electron transfer.
In Sec. \ref{isolatedcell}, we specialize the model to the case of an isolated cell.
In Sec. \ref{celllifetime}, we present an application of the proposed framework to compute the lifetime of an isolated cell.
In Sec. \ref{estimation}, we present a parameter estimation framework and fit the parameters of the model to experimental data.
Finally, Sec. \ref{future} presents some future work and Sec. \ref{conclu} concludes the paper.
\section{Stochastic cell model}
\label{stochmodel}
\begin{figure}[t]
\centering
\includegraphics[width = \linewidth,trim=20 10 15 5,clip=true]{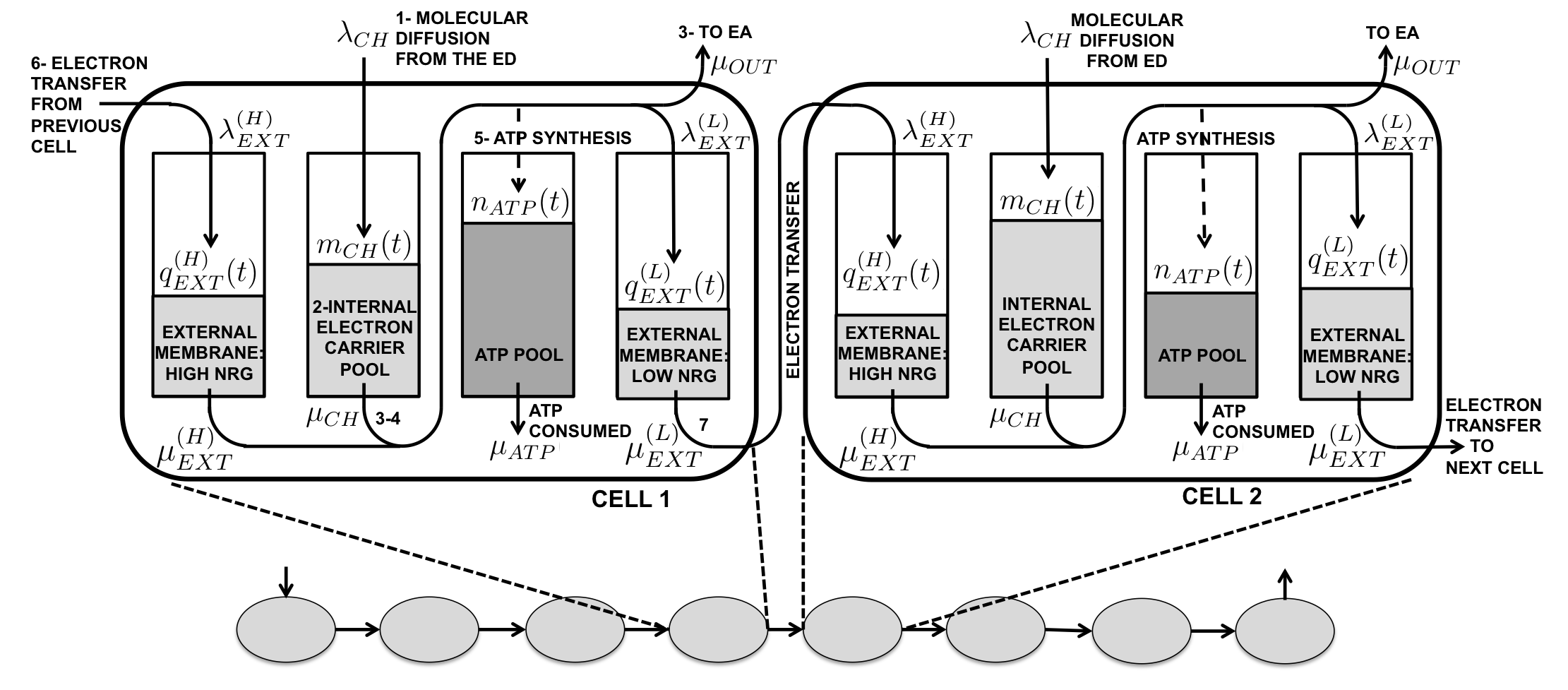}
\caption{Stochastic model of electron transfer within a bacterial cable.\vspace{-4mm}}
\label{fig1}
\vspace{-5mm}
\end{figure}
In this section, we describe a continuous time stochastic model for the dynamics of electron flow and ATP production and consumption within a single cell,
 represented in Fig.~\ref{fig1}. This is the building block of more complex multi-cellular systems, \emph{e.g.}, a bacterial cable, also represented in Fig.~\ref{fig1}.
The cell is modeled as a system with an input electron flow coming either from the ED via molecular diffusion, or from a neighboring cell via electron transfer, and an output  flow of electrons leaving either toward the EA via molecular diffusion, or toward the next cell in the cable via electron transfer.
We first review these well known biological and physical mechanisms and then provide our new stochastic model.
   Inside the cell, the conventional pathway of electron flow, enabled by the presence of the ED and the EA, is as follows (see the numbers in Fig. \ref{fig1}): 
\begin{enumerate}
\item ED molecules permeate inside the cell via molecular diffusion;
\item The presence of these ED molecules inside the cell results in reactions that produce electron-containing carriers (\emph{e.g.}, NADH).
These are collected in the \emph{internal electron carrier pool} (IECP, Fig.~\ref{fig1}).
 The electron carriers diffusively transfer electrons to the ETC, which is partially localized in the cell inner membrane;
\item The electrons originating from the electron carriers flow through the ETC and are discarded by either a soluble and internalized EA (\emph{e.g.}, molecular Oxygen) or are transferred through the periplasm to the outer membrane and deposited on an extracellular EA;
\item The electron flow through the ETC results in the production of a proton concentration gradient (proton motive force~\cite{Lane}) across the inner membrane of the cell;
\item The proton motive force is utilized by an inner membrane protein called ATP synthase to produce ATP as an energy reserve that will later be used for various functions in the cell. The ATP produced in this way is collected in the \emph{ATP pool} in Fig.~\ref{fig1}, and used by the cell
to sustain its vital operations and functions.
\end{enumerate}
Alternatively, when the cells are organized in multi-cellular structures, \emph{e.g.}, bacterial cables,
an additional pathway of electron flow may exist, termed \emph{intercellular electron transfer} (IET), which involves only a transfer of electrons
between neighboring cells, as opposed to molecules (ED and EA) diffusing through the cell membrane. In this regime, one or both of the ED and the EA are replaced by neighboring cells in a network of interconnected cells. In other words, IET can be substituted for the ED or the EA, enabling cells to survive even in the absence of the ED or the EA. In this case, the pathway for the electrons is as follows:
\begin{enumerate}
\setcounter{enumi}{5}
\item High-energy\footnote{Note that the terms \emph{high} and \emph{low} referred to the energy of electrons
 are used here only
in relative terms, \emph{i.e.}, relative to the redox potential at the  cell surface.
In bacterial cables, the redox potential slowly decreases along the cable, thus inducing a net flow of electrons from one end to the opposite one.
} electrons localized in the outer-membrane of a neighboring cell are transported to the host cell, and utilized in its ETC to produce ATP. Therefore, the electrons creating the proton motive force are not originating from the chemical carriers such as NADH, but instead are entering directly from the neighboring cell;
\item The electrons subsequently leave the ETC and move to the outer-membrane of the host cell, and are transferred to another neighboring cell that, in turn, uses these electrons to produce ATP. 
\end{enumerate}
As a result, this cooperative strategy creates a multi-cellular ETC that utilizes IET to distribute electrons throughout an entire bacterial network. These electrons originate from the ED localized on one end of the network to the locally available EA on the other end. The collective electron transport through this network provides energy for all cells involved to maintain their vital operations.
The conventional ED-EA and IET processes may coexist, depending on the availability of both ED and EA in the medium where the cell is operating
and on the connectivity of the cell to neighboring ones.
 For instance, if the concentration of ED and EA is sufficiently large, only the conventional pathway is used by the cell for ATP production.
In contrast, if such concentration is too small to support ATP production, only IET from/to neighboring cells may be active.
In accordance with the steps outlined above, we
propose the following stochastic model for the cell, as depicted in Fig.~\ref{fig1}.
 This model incorporates four pools:
\begin{enumerate}
\item The \emph{IECP}, containing the electron carrier molecules (\emph{e.g.}, NADH) produced as a result of ED
diffusion across the cell membrane and chemical processes occurring inside the cell;
\item The \emph{ATP pool}, containing all the ATP molecules produced as a result of electron flow from the electron carriers through the ETC to the EA;
\item The \emph{external membrane pool}, which involves the extracellular respiratory pathway of the cell in the outer membrane. This part of the 
ETC typically includes heme-containing c-type cytochromes that facilitate electron transfer outside of the inner membrane and into the terminal EA. In fact, the accumulation of these c-type cytochromes in the outer membrane forms the external membrane pool.
In order to incorporate the case of IET into this model, we assume that the external membrane pool is further divided into two~parts:
\begin{enumerate}
\item \emph{High energy external membrane} (HEEM), which contains high energy electrons coming from previous cells in the cable;
\item \emph{Low energy external membrane}  (LEEM), which collects low energy electrons that have been used to synthesize ATP, before they are transferred to a neighboring cell.
\end{enumerate}
\end{enumerate}
Each pool in this model has a corresponding inflow and outflow of electrons that connect that pool to the others, and one cell to the next in the cable:
\begin{enumerate}
\item The IECP gains electrons from ED molecules diffusing into the cell and transforming into electron carriers through a series of reactions;
we model this as a flow with rate $\lambda_{CH}$ joining the IECP in Fig.~\ref{fig1}. The electrons leave this pool to the ETC (cell inner membrane) to produce ATP, modeled as another flow  with rate $\mu_{CH}$ leaving the IECP in Fig.~\ref{fig1};
\item Alternatively, electrons are transferred from neighboring cells into the HEEM,
corresponding to the flow with rate $\lambda_{EXT}^{(H)}$ in Fig. \ref{fig1}. These electrons leave this pool to the ETC (cell inner membrane) to produce ATP, modeled as another flow with rate $\mu_{EXT}^{(H)}$ leaving the IECP in Fig.~\ref{fig1};
\item The electron flow out of the first pool (either the IECP or the HEEM) directly causes the synthesis of ATP, so that the overall flow into the ATP pool
 is  $\mu_{CH}+\mu_{EXT}^{(H)}$. On the other hand, ATP consumption via ATP hydrolysis within the cell through various functions is responsible for the ATP molecules leaving the ATP pool, with rate $\mu_{ATP}$;
\item As a simplification, we assume that there are two major pathways for the electron output of the ETC: internalized molecular Oxygen in aerobic conditions and transport to the external membrane in anaerobic conditions. The former case, modeled as  a flow with rate $\mu_{OUT}$ leaving the cell to the EA in Fig. \ref{fig1}, does not involve the external membrane pool but only the EA.
 In contrast, the latter involves the extracellular respiration pathway, which includes the external membrane.
 The electrons in this case are collected in the LEEM, \emph{i.e.}, the flow with rate $\lambda_{EXT}^{(L)}$ in Fig. \ref{fig1}.
 The electrons in this pool can, in turn, be transferred to neighboring cells, modeled as a flow with rate $\mu_{EXT}^{(L)}$ leaving the LEEM of cell 1 to the HEEM of cell 2 in Fig.~\ref{fig1}, or to solid phase terminal EAs, not represented in~Fig.~\ref{fig1}.
\end{enumerate}

In addition, because typical values for transfer rates between electron carriers (\emph{e.g.}, outer-membrane cytochromes) on the cell exterior are relatively high \cite{Pirbadian}, one can assume that the external membranes of neighboring cells have high transfer rates between one another, \emph{i.e.}, 
when IET is active, we have 
$\mu_{EXT}^{(L)}=\lambda_{EXT}^{(H)}=\infty$, so that any electron collected in the LEEM is instantaneously transferred to the  
HEEM of the neighboring cell in the cable. Under these assumptions, we can simplify the model
by combining the LEEM and HEEM pools of Fig. \ref{fig1} together,
so that any pair of neighboring cells share a single pool for IET.
 On the other hand, if the cell is isolated, no IET occurs, hence
$\mu_{EXT}^{(L)}=0$ and/or $\lambda_{EXT}^{(H)}=0$. This latter case will be studied in more detail in Sec. \ref{isolatedcell}.
   
We model the cell as a finite state machine, and characterize the state of the cell and its stochastic evolution.
The \emph{internal state} of a given cell at time $t$ is defined as
\begin{align}\label{Si}
\mathbf s_I(t)=\left(m_{CH}(t),n_{ATP}(t),q_{EXT}^{(L)}(t),q_{EXT}^{(H)}(t)\right),
\end{align}
where:
\begin{itemize}
\item $m_{CH}(t)$ is the number of electrons in the IECP that will participate in the synthesis of ATP; these electrons are carried
by ED units which diffuse through the membrane into the cell (\emph{e.g.}, lactate), and are bonded to electron carriers within the cell (\emph{e.g.}, NADH);
$m_{CH}(t)$ takes value in the set $\mathcal M_{CH}\equiv\{0,1,\dots,M_{CH}\}$,
 where $M_{CH}$ is the \emph{electro-chemical storage capacity} of the cell;
\item $n_{ATP}(t)$ is the number of ATP molecules within the cell, taking value in the set $\mathcal N_{AXP}\equiv\{0,1,\dots,N_{AXP}\}$,
 where $N_{AXP}$ is the 
overall number of ATP plus ADP molecules in the cell, which is
assumed to be constant over time;
 $N_{AXP}$ also represents
the maximum number of ATP molecules which can be present within the cell at any time (when no ADP is present);
\item $q_{EXT}^{(H)}(t)$ is the number of electrons in the HEEM, taking value in the set $\mathcal Q_{EXT}^{(H)}\equiv \{0,1,\dots,Q_{EXT}^{(H)}\}$,
where $Q_{EXT}^{(H)}$ is the electron "storage capacity" of the HEEM;
\item $q_{EXT}^{(L)}(t)$ is the number of electrons in the LEEM, taking value in the set $\mathcal Q_{EXT}^{(L)}\equiv \{0,1,\dots,Q_{EXT}^{(L)}\}$,
where $Q_{EXT}^{(L)}$ is the electron "storage capacity" of the LEEM.
\end{itemize}
\begin{remark}
For simplicity, we assume that
all the quantities related to the state of the cell and to the flows of electrons/molecules are in terms of equivalent number of electrons involved, rather than molecular units. Hence, for instance,
the ATP level in the ATP pool, $n_{ATP}(t)$, actually represents the equivalent number of electrons involved in the synthesis
of the corresponding quantity of ATP available in the cell.
Similarly, the level of NADH in the IECP, $m_{CH}(t)$, is expressed in terms of the equivalent number of electrons carried
by the electron carriers, which actively synthesize ATP.
A similar interpretation holds for the flows (of electrons, rather than molecules  or $\meas{mM}$, where $\meas{M}$ stands for "1 molar").
Transition from one representation (electrons) to the other (molecules or $\meas{mM}$) is possible by appropriate scaling.

Moreover, while in the following analysis we assume that one "unit" corresponds to one electron, this can be generalized to the case
where one "unit" corresponds to $N_E$ electrons, so that, \emph{e.g.}, $n_{ATP}$ units in the ATP pool correspond to
$N_E n_{ATP}$ electrons.
\end{remark}
Note that, if the cell is connected to other cells in a larger community, the low (respectively, high) energy external membrane is shared with the high (low) energy external membrane of the neighboring cell, owing to the \emph{high transfer rate approximation}, as explained above.
  Additionally, we denote the state of death of the cell as $\text{DEAD}$ (to be specified later).
  The state space of the cell is denoted as
\begin{align*}
\mathcal S_I\equiv \left(\mathcal M_{CH}\times\mathcal N_{AXP}\times\mathcal Q_{EXT}^{(L)}\times\mathcal Q_{EXT}^{(H)}\right)\cup\{\text{DEAD}\}.
\end{align*}
Note that the behavior of the cell is influenced by the concentration of the ED and the EA in the surrounding medium.
Therefore, we also define the \emph{external state} of the cell as $\mathbf s_{E}(t)=(\sigma_{D}(t),\sigma_{A}(t))$,
  where $\sigma_D(t)$ and $\sigma_A(t)$ are, respectively, the external concentration of the ED and the EA.
  For simplicity, we assume that $\mathbf s_{E}(t)$ is an exogenous process, not influenced by the cell dynamics, \emph{i.e.}, 
  the  consumption of the ED and the EA by the cell does not influence their concentration in the surrounding medium. This requires that the medium in which cells are suspended is continuously being replaced by fresh medium containing a constant amount of the ED and the EA. Otherwise a high cell concentration would use up all the resources in the time-scales relevant to this model. This aspect will be considered in future work, and is beyond the scope of the current paper.
 \begin{figure*}
\begin{center}
\setlength{\tabcolsep}{1mm}
\begin{tabular}{cccc}
\subfigure[\emph{ED diffusion}: One electron
is transported by the ED through the cell membrane 
and is collected in the IECP]{\includegraphics*[width=0.24\linewidth]{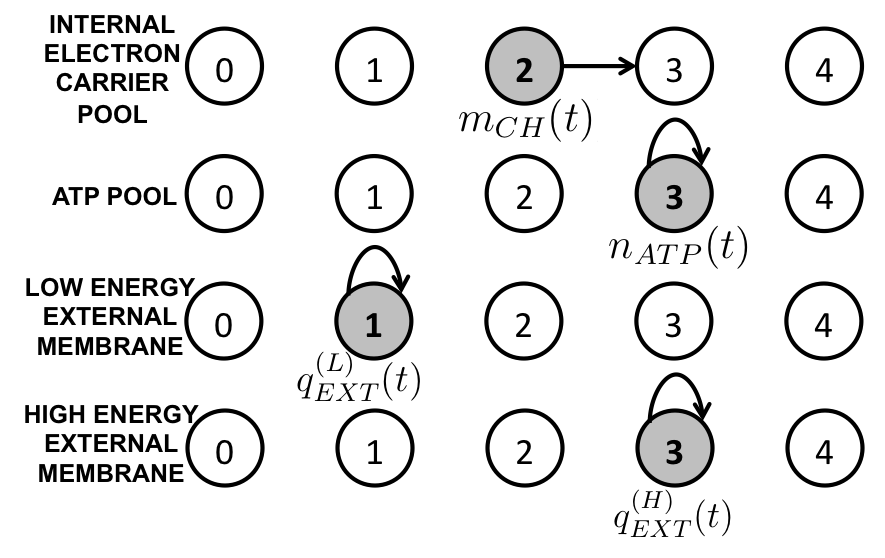}}&
\subfigure[\emph{IET}: 
One electron
is collected in the HEEM via IET from a neighboring cell]{\includegraphics*[width=0.24\linewidth]{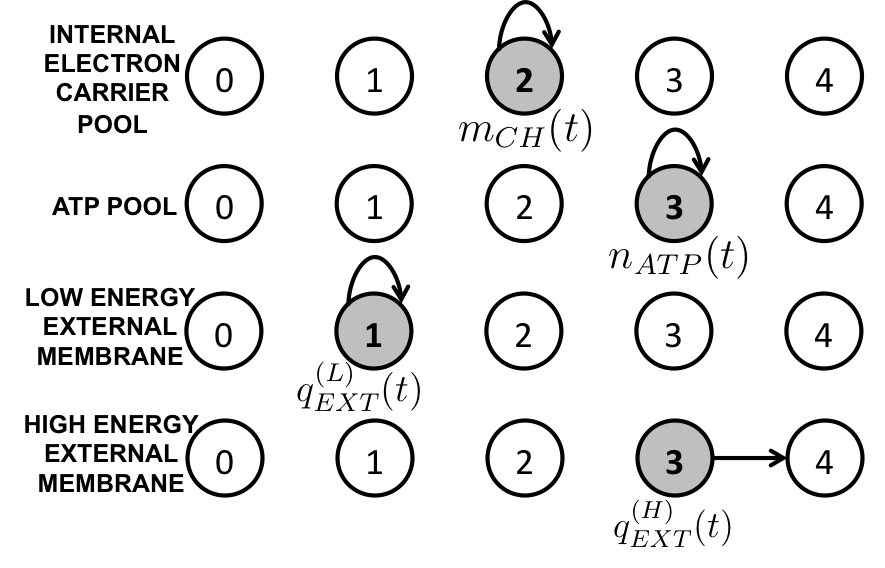}}
&
\subfigure[\emph{Conventional aerobic ATP synthesis}: One electron
is taken from the IECP to synthesize ATP, and is then captured by an EA, leaving the cell]{\includegraphics*[width=0.24\linewidth]{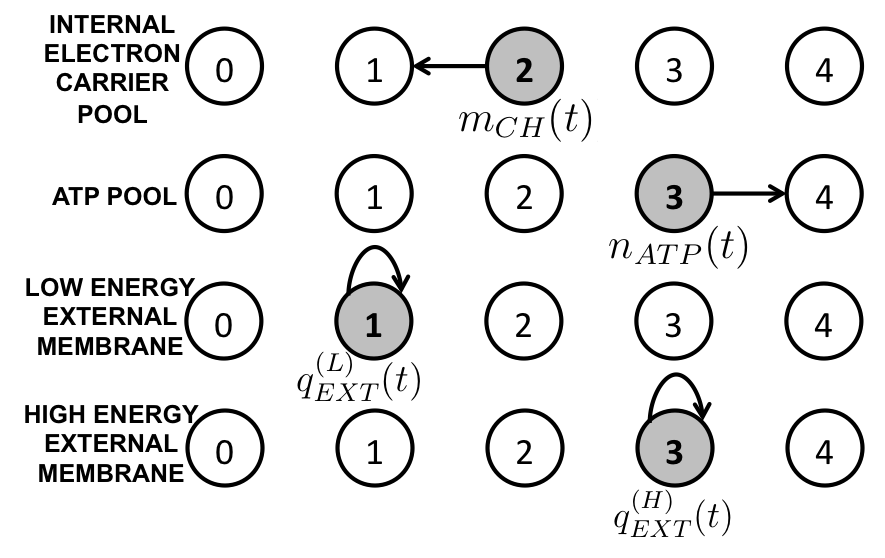}}&
\subfigure[\emph{Conventional anaerobic ATP synthesis}:  One electron
is taken from the IECP to synthesize ATP, and is  then collected in the LEEM]{\includegraphics*[width=0.24\linewidth]{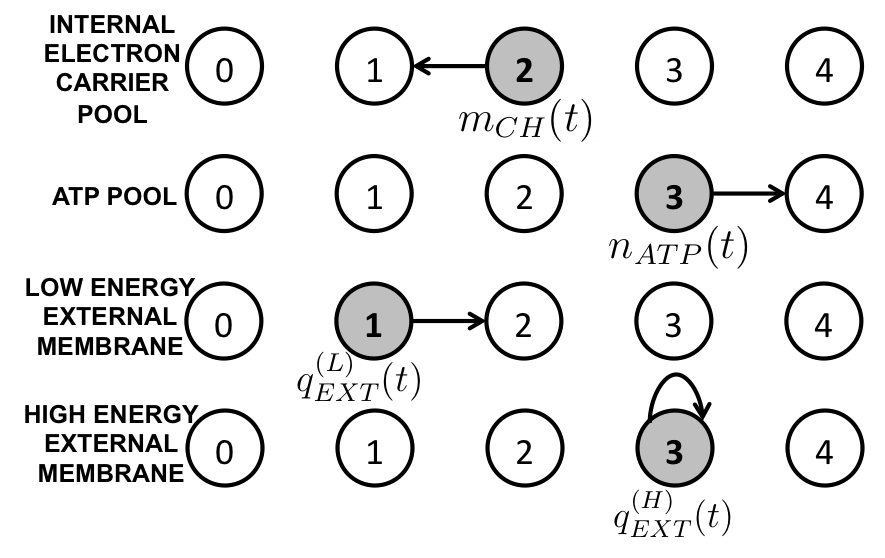}}\\
\subfigure[\emph{Unconventional aerobic ATP synthesis}:  One electron
is taken from the HEEM to synthesize ATP, and is then captured by an EA]{\includegraphics*[width=0.24\linewidth]{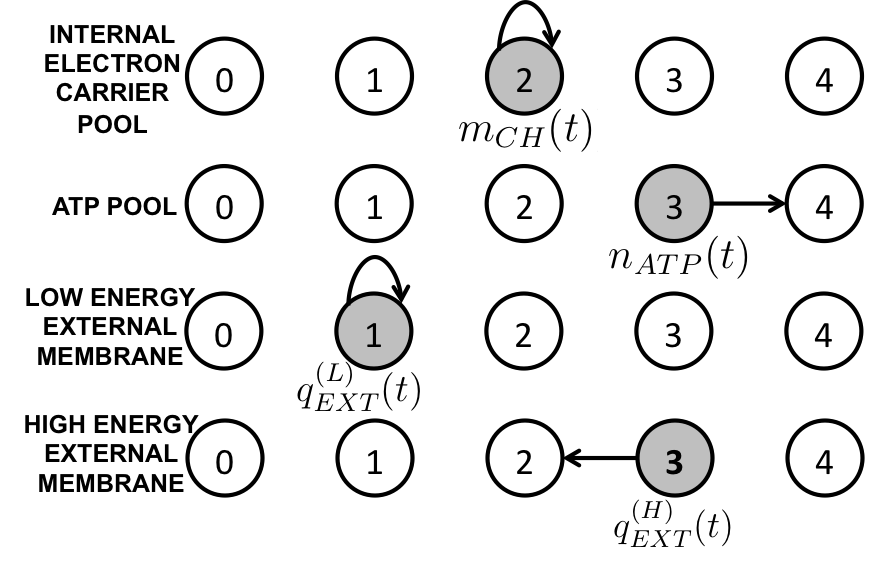}}&
\subfigure[\emph{Unconventional anaerobic ATP synthesis}: One electron
is taken from the HEEM to synthesize ATP, and is then collected in the LEEM]{\includegraphics*[width=0.24\linewidth]{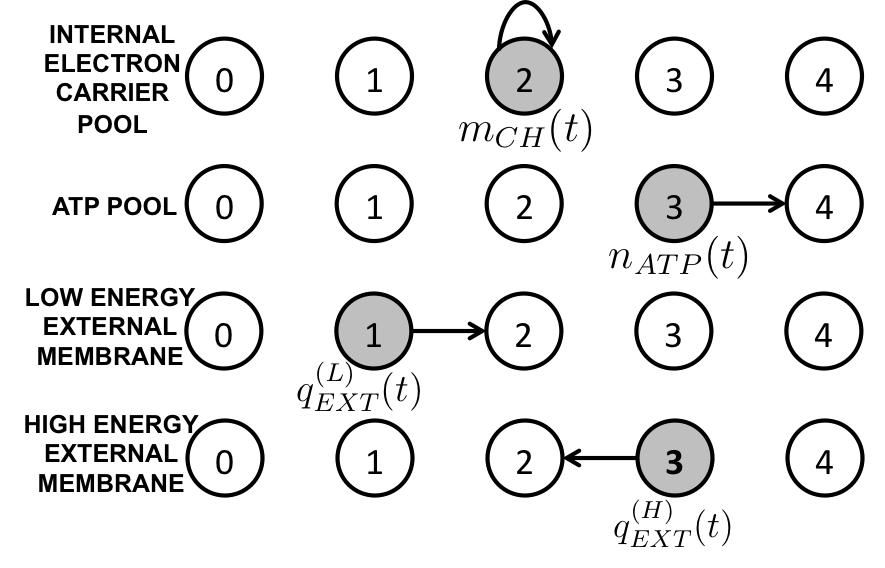}}
&\subfigure[\emph{ATP consumption}: One ATP molecule is consumed to produce energy for the cell]{\includegraphics*[width=0.24\linewidth]{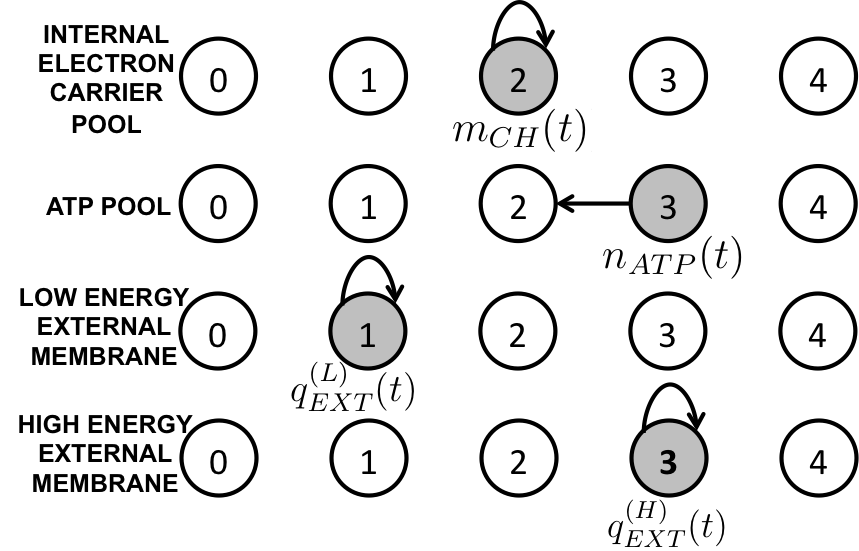}}
\end{tabular}
\caption{Markov chain and transitions from state $\mathbf s_I(t)=(2,3,1,3)$, \emph{i.e.}, two electrons are in the IECP,
 three ATP units are in the ATP pool, one electron is in the LEEM, and three electrons are in the HEEM, respectively (see Fig. \ref{fig1})\vspace{-4mm}}
\label{fig3}
\end{center}
\end{figure*}

The internal state process of cell $i$,  $\mathbf s_I^{(i)}(t)\in\mathcal S_I$ (see Eq. \ref{Si}), is time-varying and stochastic; $\mathbf s_I^{(i)}(t)$ evolves as a consequence of electro/chemical reactions occurring within the cell,
 chemical diffusion through the cell membrane, and IET from the neighboring cell $i-1$ to the neighboring cell $i+1$.
 The evolution of $\mathbf s_I^{(i)}(t)$ is also influenced by the external state $\mathbf s_E^{(i)}(t)$ experienced by the cell.
 We define the following processes affecting the evolution of $\mathbf s_I^{(i)}(t)$, all of which, for analytical tractability, are modeled as Poisson processes with
 state-dependent rates; these processes are represented in Fig.~\ref{fig1} and the corresponding state transitions are depicted in Fig.~\ref{fig3}:
\begin{itemize}
\item \underline{ED diffusion} through the membrane: ED molecules carry electrons to synthesize ATP, which are stored in the 
IECP; this process occurs with rate
$\lambda_{CH}(\mathbf s_I^{(i)}(t);\mathbf s_{E}^{(i)}(t))$ [electrons/s].
Whenever an ED diffuses through the membrane within the cell (say, at time $t$), the state
$m_{CH}^{(i)}(t)$ increases by one unit (Fig.~\ref{fig3}.a),
so that the internal state moves from $\mathbf s_I^{(i)}(t)=(m_{CH},n_{ATP},q_{EXT}^{(L)},q_{EXT}^{(H)})$ at time $t$ to $\mathbf s_I^{(i)}(t^+)=(m_{CH}+1,n_{ATP},q_{EXT},q_{EXT})$ at time instant~$t^+$;
\item \underline{IET} from the neighboring cell $i-1$: the electron is collected in the HEEM,
so that the corresponding state increases by one unit and
$\mathbf s_I^{(i)}(t^+)=(m_{CH},n_{ATP},q_{EXT}^{(L)},q_{EXT}^{(H)}+1)$ (Fig.~\ref{fig3}.b);
note that this process is coupled with the \emph{anaerobic ATP synthesis} 
(see definition below) process of the neighboring cell $i-1$ from which the electron is transferred;
in fact, owing to the high transfer rate approximation, the LEEM of cell $i-1$ is shared with the HEEM of cell $i$,
so that the rate of electron flow into the HEEM of cell $i$ is $\lambda_{EXT}^{(L)}(\mathbf s_I^{(i-1)}(t);\mathbf s_E^{(i-1)}(t))$;
\item \underline{Conventional ATP synthesis}: this process involves the transfer of one electron from the IECP
to the internal membrane to synthesize ATP, with rate $\mu_{CH}(\mathbf s_I^{(i)}(t),\mathbf s_E^{(i)}(t))$ [electrons/s].
Correspondingly, one molecule of ATP is generated; the electron then leaves the internal membrane
and follows either the aerobic pathway (\emph{i.e.}, it is captured by an internalized EA, such as Oxygen, see Fig.~\ref{fig3}.c), with overall rate
$\mu_{OUT}(\mathbf s_I^{(i)}(t);\mathbf s_E^{(i)}(t))$,
or the anaerobic one (Fig.~\ref{fig3}.d) and is collected in the LEEM, with overall rate $\lambda_{EXT}(\mathbf s_I^{(i)}(t);\mathbf s_E^{(i)}(t))$
(note that this is also the HEEM of cell $i+1$).
If the aerobic pathway is followed, the new state becomes
 $\mathbf s_I^{(i)}(t^+)=(m_{CH}-1,n_{ATP}+1,q_{EXT}^{(L)},q_{EXT}^{(H)})$ (Fig.~\ref{fig3}.c).
Otherwise (anaerobic pathway), the new state becomes
 $\mathbf s_I^{(i)}(t^+)=(m_{CH}-1,n_{ATP}+1,q_{EXT}^{(L)}+1,q_{EXT}^{(H)})$ (Fig.~\ref{fig3}.d);
  \item \underline{Unconventional ATP synthesis}: this process involves the transfer of one electron from the 
  HEEM to the internal membrane to synthesize ATP, with rate $\mu_{EXT}^{(H)}(\mathbf s_I^{(i)}(t),\mathbf s_E^{(i)}(t))$ [electrons/s].
  Afterwards, the electron follows a similar path as in the conventional ATP synthesis, \emph{i.e.}, either
it is captured by an internalized EA (aerobic pathway), with overall rate $\mu_{OUT}(\mathbf s_I^{(i)}(t);\mathbf s_E^{(i)}(t))$,
or it is collected in the LEEM of the cell (anaerobic pathway), with overall rate $\lambda_{EXT}^{(L)}(\mathbf s_I^{(i)}(t);\mathbf s_E^{(i)}(t))$.
In the former case, the new state becomes
 $\mathbf s_I^{(i)}(t^+){=}(m_{CH},n_{ATP}{+}1,q_{EXT}^{(L)},q_{EXT}^{(H)}{-}1)$ (Fig.~\ref{fig3}.e);
 in the latter, $\mathbf s_I^{(i)}(t^+){=}(m_{CH},n_{ATP}{+}1,q_{EXT}^{(L)}{+}1,q_{EXT}^{(H)}-1)$ (Fig.~\ref{fig3}.f);
\item \underline{ATP consumption}: this process provides energy for cellular functions, and occurs with rate $\mu_{ATP}(\mathbf s_I^{(i)}(t);\mathbf s_E^{(i)}(t))$ [electrons/s];
when one molecule of ATP is consumed, the state $n_{ATP}^{(i)}(t)$ decreases by one unit,
so that $\mathbf s_I^{(i)}(t^+)=(m_{CH},n_{ATP}-1,q_{EXT}^{(L)},q_{EXT}^{(H)})$ (Fig.~\ref{fig3}.g);
\item \underline{Death} process, with rate $\delta(\mathbf s_I^{(i)}(t);\mathbf s_E^{(i)}(t))$:
if death occurs, the new state becomes $\mathbf s_I^{(i)}(t^+)=\text{DEAD}$,
from which the cell cannot recover any longer, \emph{i.e.},
$\mathbf s_I^{(i)}(\tau)=~\text{DEAD},\ \forall\tau>~t$.
\end{itemize}
\subsection{Flow Constraints}\label{constr}
Note that the rates of the different flows involved need to satisfy some constraints, induced by the queuing model employed.
In particular, if some queue is empty (respectively, saturated), the rate of the corresponding outbound (respectively, inbound) flow must be zero,
so that, for instance, for the flows out of and into the ATP pool, the following condition must hold:
\begin{align*}
&\mu_{ATP}(m_{CH},0,q_{EXT}^{(L)},q_{EXT}^{(H)};\mathbf s_E)=0, \ \text{(outbound flow)},\\
&\mu_{CH}(m_{CH},N_{AXP},q_{EXT}^{(L)},q_{EXT}^{(H)};\mathbf s_E)\\&
+\mu_{EXT}^{(H)}(m_{CH},N_{AXP},q_{EXT}^{(L)},q_{EXT}^{(H)};\mathbf s_E){=}0, \ \text{(inbound flow)}.
\end{align*}
A similar consideration holds for the other queues and the corresponding flows.
Moreover,
\begin{align*}
&\mu_{EXT}^{(H)}(\mathbf s_I;\mathbf s_E){+}\mu_{CH}(\mathbf s_I;\mathbf s_E)
{=}
\lambda_{EXT}^{(L)}(\mathbf s_I;\mathbf s_E){+}\mu_{OUT}(\mathbf s_I;\mathbf s_E),
\end{align*}
since each electron leaving either the IECP or the HEEM
to synthesize ATP either follows the aerobic pathway to the EA or the anaerobic one to the LEEM.

We further assume that
\begin{align}
&\lambda_{CH}(\mathbf s_I;\sigma_D,\sigma_A)=\sigma_D\lambda_{CH}(\mathbf s_I;1,\sigma_A),
\\
&\mu_{OUT}(\mathbf s_I;\sigma_D,\sigma_A)=\sigma_A\mu_{OUT}(\mathbf s_I;\sigma_D,1),
\end{align}
thus capturing the fact that 
 the molecular diffusion rate is proportional to the ED (respectively, EA) concentration. This assumption is supported by Fick's law of diffusion \cite{smith2003foundations}, which states that the diffusion rate is linearly dependent on the concentration differential between inside and outside.
It follows that, if no ED is present ($\sigma_D=0$), then $\lambda_{CH}(\mathbf s_I;0,\sigma_A)=0$
and no ED diffusion may occur.
Similarly, if no EA is present ($\sigma_A=0$), then $\mu_{OUT}(\mathbf s_I;\sigma_D,0)=0$ and no EA diffusion may occur.
In Sec. \ref{parammodel}, a parametric model for these flows is presented, based on which the model is fit to experimental~data.
\section{Isolated cell model}\label{isolatedcell}
In the most general case, electron transport in a series of interconnected single-cell organisms is represented by the proposed stochastic model. However, this model can also explain the electron transport behavior of a single cell, which is the building block of the general multi-cell system. The experimental investigation of a multi-cellular network of bacteria is very challenging, in fact:
\begin{enumerate}
\item In order to build a chain of interconnected cells, single-cell organisms have to be placed in each other's proximity. Placing multiple cells next or close to each other in a controlled way that maintains the intercellular contact is very difficult in practice and requires cellular manipulation techniques such as optical tweezers \cite{Liu2}, as well as nanofabricated micron-scaled chambers designed specifically to hold these communities in place;
\item \emph{In vivo} characterization of the energetic and electron transfer properties of an individual cell within this chain independently from the other cells requires complex chemical and optical assays that have never been used in such complicated systems.
\end{enumerate}
Therefore, instead of the most general case of the model (multi-cell system), we start by investigating the properties of single, isolated cells. Using a few simplifying assumptions, the general model can be reduced to a single cell model which can be more easily matched against experimental results. In addition, the single-cell experiments are not hindered by the practical issues mentioned above, which makes them easier to perform. In this way, we can characterize the properties of the individual components, which will help us better understand the electron transport in multi-cellular systems. 

\begin{figure}
\begin{center}
\setlength{\tabcolsep}{0mm}
\begin{tabular}{MM}
\subfigure{\includegraphics*[width=.95\linewidth,trim=0 5 0 5,clip=true]{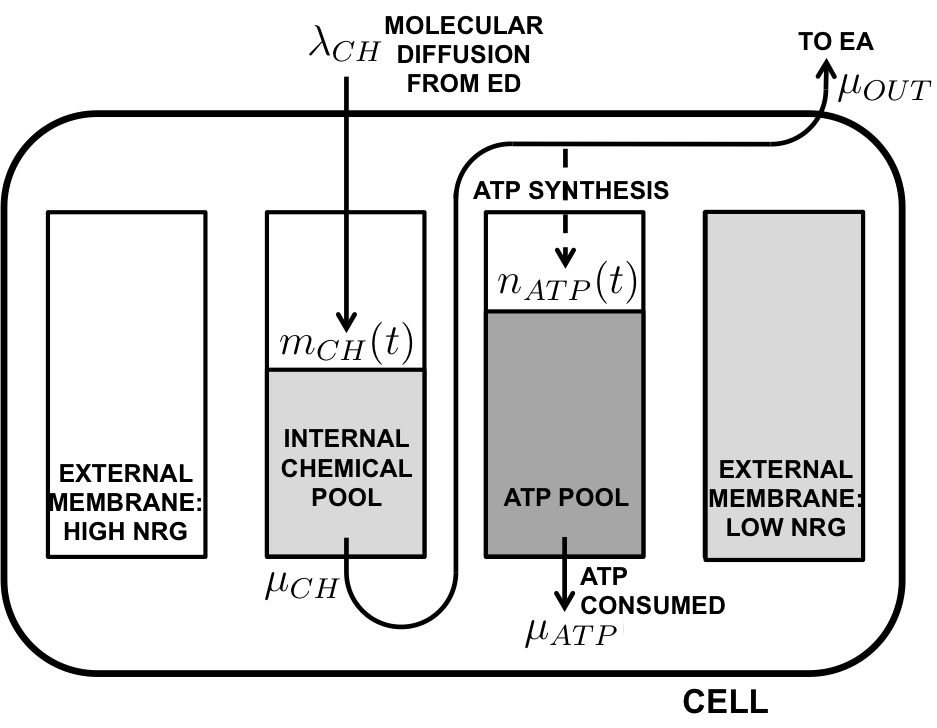}}&
\subfigure{\includegraphics*[width=.95\linewidth]{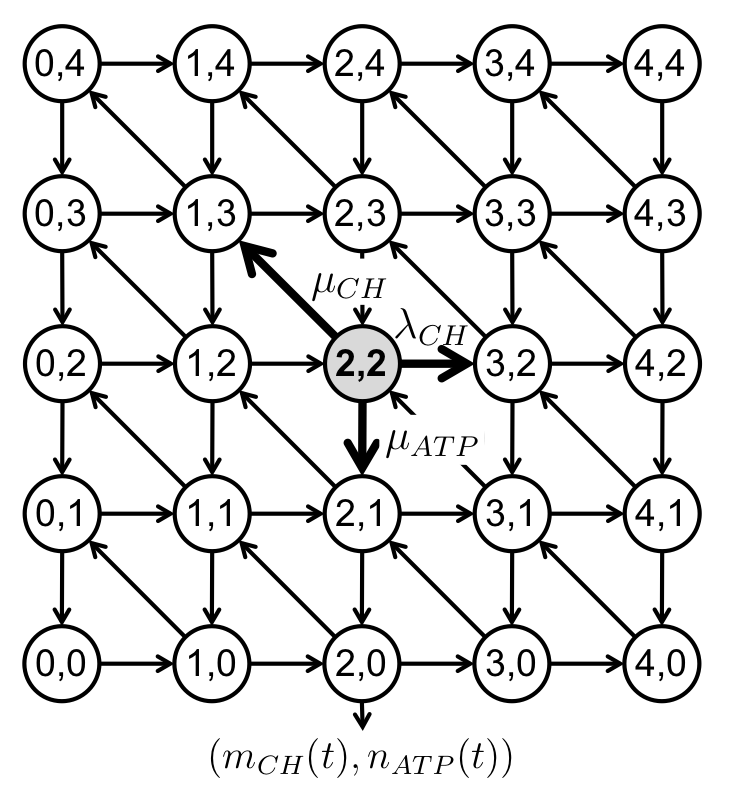}}
\end{tabular}
\caption{
Stochastic model for an isolated cell, after the transient phase during which the HEEM gets depleted and the LEEM gets charged (left),
and
Markov chain with the corresponding transitions (right), for the case where $M_{CH}=4$, $N_{AXP}=4$.
The transition rates from state $(2,2)$ are also depicted.\vspace{-4mm}}
\label{fig5}
\end{center}
\vspace{-3mm}
\end{figure}

In the case of an isolated cell, the IET process is not active, and
$\lambda_{EXT}^{(H)}(t)=\mu_{EXT}^{(L)}(t)=0$.
As a result, the HEEM gets depleted, and the LEEM gets filled.
Therefore, after a transient phase, the cell reaches the configuration depicted in Fig. \ref{fig5},
where the HEEM is empty, and the LEEM is fully charged.
 In the following treatment, we assume that the transient phase is concluded, hence $q_{EXT}^{(L)}(t)=Q_{EXT}^{(L)}$ and $q_{EXT}^{(H)}(t)=0,\ \forall t$, so that the 
state $(q_{EXT}^{(L)}(t),q_{EXT}^{(H)}(t))=(Q_{EXT}^{(L)},0)$ of the external membrane can be neglected.
Assuming that the cell operates in this configuration, we thus redefine  its internal state as $\mathbf s_I(t)=(m_{CH}(t),n_{ATP}(t))$. 

 Ê
The corresponding Markov chain and state transitions are depicted in Fig. \ref{fig5}.
From the continuous-time process described in Sec. \ref{stochmodel},
we now generate a discrete-time process, as detailed below.
Initially, we assume that the external state $\mathbf s_E(t)$ is fixed, \emph{i.e.}, $\sigma_D(t)=\sigma_D,\ \forall t$ and 
$\sigma_A(t)=\sigma_A,\ \forall t$.
The case where $\mathbf s_E(t)$ is piecewise constant will be considered in Sec. \ref{senonconstant}.
The discretization is obtained by sampling the state process $\mathbf s_I(t)$ at specific times,
corresponding to one of the events described in Sec. \ref{stochmodel}, specialized to the case of an isolated cell: molecular diffusion; conventional aerobic ATP synthesis; ATP consumption; death.
Starting from time $t=0$ in state $\mathbf s_I(0)\in\mathcal S_I$, we define
$T_k$ as the time instant corresponding to the occurrence of the $k$th event since time $0$, and $\mathbf S_k$ as the corresponding state at time instant $T_k^+$
(\emph{i.e.}, right after the corresponding transition occurs). In particular, $T_0=0$ and $\mathbf S_0=\mathbf s_I(0)$.
Note that, by sampling, we have transformed the continuous-time stochastic process into a
discrete-time Markov chain, with finite state space $\mathcal S_I$.
However, the duration of the $k$th time-slot, $T_{k+1}-T_{k}$, is not fixed but is a random variable which depends on the inter-arrival time of the events described in Sec. \ref{senonconstant}.
In the subsequent sections, we first derive the transition probabilities of the underlying discrete time Markov chain and the inter-arrival times of the events,
thus leading to a full-characterization of the stochastic dynamics of $\mathbf s_I(t)$.
We then provide an example of applicability of this framework to the computation of the lifetime of the cell.
Finally, in Sec. \ref{estimation}, we present a parameter estimation framework and match the model to experimental data available in \cite{Ozalp}. 
\subsection{Transition Probabilities and inter-arrival times}
\label{TXprob}
In this section, we compute the transition probability of the underlying discrete-time Markov chain, and the distribution of the inter-arrival times 
in the corresponding continuous time system.
To this end, let $\mathbf S_k=i\footnote{In this section, $i$ is an index corresponding to a specific state in $\mathcal S_I$.}\in\mathcal S_I\setminus\{\text{DEAD}\}$
 be the state of the cell at time $T_k^+$.
We compute the transition probability
\begin{align}\label{prob}
\mathbb P(\mathbf S_{k+1}=j,T_{k+1}>\tau |\mathbf S_k=i,T_k=t),
\end{align}
for some $j\in\mathcal S_I$, $\tau\geq t$
(note that, due to the memoryless property of Poisson processes, the event $\mathbf S_{k+1}=j,T_{k+1}>\tau $ conditioned on
$\mathbf S_k=i,T_k=t$ is independent of the realization of $\{(\mathbf S_j,T_j),0\leq j<k\}$).
Let $\lambda_{i,j}$ be the transition rate from state $i$ to state $j$, which depends on the specific event
which triggers the transition. For instance, if $i$ corresponds to $(m_{CH},n_{ATP})$ and $j$ to $(m_{CH},n_{ATP}-1)$, then a transition from state $i$ to state $j$ occurs if the ATP consumption event occurs, with rate $\lambda_{i,j}=\mu_{ATP}(\mathbf s_I;\mathbf s_E)$.
The transition from state $i$ to state $j$ can be interpreted as follows. Let $E_{i,s}$ be the event which triggers
the transition from $i$ to $s$, and $t+W_{i,s}$ be the time when such event occurs (with respect to the reference time-position $t$).
From the properties of Poisson processes, we have that
$W_{i,s}$ is an exponential random variable, with pdf $f_{W_{i,s}}(w)=\lambda_{i,s}e^{-\lambda_{i,s}w}$,
and that $\{W_{i,s},\forall s\}$ are mutually independent.
Then, the system moves to state $j$ if $t+W_{i,j}<t+W_{i,s},\ \forall s\neq j$, \emph{i.e.}, the event $E_{i,j}$ is the first one to occur, which thus triggers the transition. Therefore,
the probability (\ref{prob}) is equivalent to
\begin{align}
\label{p1}
&\mathbb P(\mathbf S_{k+1}=j,T_{k+1}>\tau |\mathbf S_k=i,T_k=t)
\\&
=
\mathbb P(t+W_{i,j}>\tau,W_{i,j}<W_{i,s},\ \forall s\neq j |\mathbf S_k=i,T_k=t)
\nonumber\\&
=
\int_{\tau-t}^\infty
\lambda_{i,j}e^{-\lambda_{i,j}w}
\prod_{s\neq j}\mathbb P(W_{i,s}>w)\mathrm dw
=
\frac{\lambda_{i,j}}{R_i}e^{-R_i(\tau-t)},\nn
\end{align}
where we have
defined the total flow from state $i$, $R_i=\sum_{s}\lambda_{i,s}$,
we have marginalized
with respect to $W_{i,j}$, we have used the independence
among $\{W_{i,s},\forall s\}$ and $\mathbb P(W_{i,s}>w)=e^{-\lambda_{i,s}w}$.
From (\ref{p1}), we thus obtain the transition probability
$\mathbb P(\mathbf S_{k+1}=j|\mathbf S_k=i)$ by letting $\tau=t$ in (\ref{p1}) and by noticing that the resulting expression is independent of $t$, \emph{i.e.},
\begin{align*}
&\mathbb P(\mathbf S_{k+1}=j|\mathbf S_k=i,T_k=t)=
\frac{\lambda_{i,j}}{R_i}
=\mathbb P(\mathbf S_{k+1}=j|\mathbf S_k=i).
\end{align*}
We now compute the distribution of the inter-arrival time $T_{k+1}-T_{k}$ as
\begin{align*}
&\mathbb P(T_{k+1}-T_k>\tau-t |\mathbf S_k=i,\mathbf S_{k+1}=j,T_k=t)
\nn\\&=
\frac{\mathbb P(\mathbf S_{k+1}=j,T_{k+1}>\tau |\mathbf S_k=i,T_k=t)}
{\mathbb P(\mathbf S_{k+1}=j|\mathbf S_k=i,T_k=t)}
=e^{-R_i(\tau-t)}.
\end{align*}
Note that the resulting expression is independent of $\mathbf S_{k+1}$ and of time $t$, since the process is stationary.
We can thus write
\begin{align}
&\mathbb P(T_{k+1}-T_k>\tau-t |\mathbf S_k=i)=e^{-R_i(\tau-t)}.
\end{align}

We define the $(|\mathcal S_I|-1)\times (|\mathcal S_I|-1)$ transition probability matrix $\mathbf T$ of the underlying discrete-time Markov chain within $\mathcal S_I\setminus\{\text{DEAD}\}$, with entries $\mathbf T(i,j)=\mathbb P(\mathbf S_{k+1}=j|\mathbf S_k=i),\ i,j\in\mathcal S_I\setminus\{\text{DEAD}\}$ (we do not consider transitions from $\text{DEAD}$, since this is absorbing).
The transition probability from $i\in\mathcal S_I\setminus\{\text{DEAD}\}$ to $\text{DEAD}$ is then given by
$1-\mathbf e_i^T\mathbf T\mathbf 1$, where $\mathbf 1$ is the column vector of all ones, and $\mathbf e_i$ 
equals $1$ in the position corresponding to state $i$, and zero otherwise.
\subsection{State distribution of the system at time $t>0$}
Given the analysis of the underlying discrete-time Markov chain and of the inter-arrival times
in the previous section, we are now able to compute the state distribution of the system at 
a generic time $t$, given that $\mathbf S_I(0)=i$.
  We define 
 \begin{align}
\mathbb P_t(j|i)=\mathbb P(\mathbf S_I(t)=j|\mathbf S_I(0)=i),\ j\in\mathcal S_I\setminus\{\text{DEAD}\}.
 \end{align}
In order to compute it, let $0<h<t$. By the memoryless property of Poisson processes,
 \begin{align}
\mathbb P_t(j|i)&=\!\!\!\!\!\sum_{s\in\mathcal S_I\setminus\{\text{DEAD}\}}\mathbb P(\mathbf S_I(t)=j,\mathbf S_I(t-h)=s|\mathbf S_I(0)=i)
\nn\\&
=\!\!\!\!\!
\sum_{s\in\mathcal S_I\setminus\{\text{DEAD}\}}\mathbb P_h(j|s)\mathbb P_{t-h}(s|i).
 \end{align}
 It follows that, $\forall i,j\in\mathcal S_I\setminus\{\text{DEAD}\}$,
  \begin{align*}
\mathbb P_t(j|i)-\mathbb P_{t-h}(j|i)
=\!\!\!\!\!
\sum_{s\in\mathcal S_I\setminus\{\text{DEAD}\}}(\mathbb P_h(j|s)-\delta_{j,s})\mathbb P_{t-h}(s|i).
 \end{align*}
  Then, dividing by $h$ and taking the limit for $h\to 0$, we obtain
   \begin{align}
   \label{p2}
   \frac{\mathrm d\mathbb  P_{t}(j|i)}{\mathrm d t}
=
\sum_{s\in\mathcal S_I\setminus\{\text{DEAD}\}}\lim_{h\to 0}\frac{\mathbb P_h(j|s)-\delta_{j,s}}{h}
\mathbb P_{t}(s|i).
 \end{align}
Note that
 $\underset{h\to 0}\lim\frac{\mathbb P_h(j|s)-\delta_{j,s}}{h}\!=\!\lambda_{s,j}$, 
 and
$\underset{h\to 0}\lim\frac{\mathbb P_h(s|s)-\delta_{s,s}}{h}\!=\!-R_s$.
Substituting in (\ref{p2}), we obtain the system of differential equations
   \begin{align}
   \label{p3}
   \frac{\mathrm d\mathbb  P_{t}(j|i)}{\mathrm d t}
=
\!\!\!\!\!\sum_{s\in\mathcal S_I\setminus\{\text{DEAD},j\}}\!\!\!\!\!\lambda_{s,j}\mathbb P_{t}(s|i)
-R_j\mathbb P_{t}(j|i),\quad\forall i,j.
 \end{align}
 Letting $\mathbf P_{t}$ be the $(|\mathcal S_I|-1)\times(|\mathcal S_I|-1)$ matrix with components
 $\mathbf P_{t}(i,j)=\mathbb P_{t}(j|i),\ i,j\in\mathcal S_I\setminus\{\text{DEAD}\}$,
 we can rewrite the system of differential equations (\ref{p3}) as
   \begin{align}
   \label{p4}
   \mathbf  P_{t}^\prime
=
\mathbf P_{t}\mathbf A,
 \end{align}
 where we have defined the \emph{flow matrix} $\mathbf A$ with components
 $\mathbf A(s,j)=\lambda_{s,j}$ for $j\neq s$ and 
 $\mathbf A(j,j)=-R_j$, and $\mathbf  P_{t}^\prime$ represents the first-order derivative of $\mathbf P_{t}$ with respect to time.
Note that  $\mathbf A=\mathbf R(\mathbf T-\mathbf I)$,
where $\mathbf T$ is the transition matrix of the underlying discrete-time Markov chain within $\mathcal S_I\setminus\{\text{DEAD}\}$, derived in the previous section,
 $\mathbf R$ is the \emph{rate matrix}, a diagonal matrix with entries $\mathbf R(i,i)=R_i$, and $\mathbf I$ is the unit matrix.
Moreover, by Gershgorin's circle Theorem \cite{Golub}, all eigenvalues of $\mathbf A$ are non-positive.
The general solution to (\ref{p4})  subject to $\mathbf P_0=\mathbf I$ is
\begin{align}
\label{gensol}
\mathbf P_t=\exp\{\mathbf At\},
\end{align}
where we have defined the matrix exponential $\exp\{\mathbf At\}=\sum_{k=0}^{\infty}\frac{t^k}{k!}\mathbf A^k$.
Note that such solution guarantees a feasible transition probability matrix, \emph{i.e.},
 $[\mathbf P_t]_{i,j}\geq 0$ and $\sum_{j}[\mathbf P_t]_{i,j}\leq~1$.
 \subsection{Numerical evaluation of $\mathbf P_t$}

  Unfortunately, from our numerical evaluations, we have verified that $\mathbf A$ can seldom be diagonalized.
  Therefore, we employ an alternative solution to efficiently compute $\mathbf P_t$. Let $\Delta\ll 1$
and $n=\lceil t/\Delta\rceil$. Then, the general solution can be approximated as
\begin{align}
\label{approx}
\!\!\!\!\mathbf P_t\!=\!\left[\exp\{\mathbf A\Delta\}\right]^{n}\!\exp\{\mathbf A(t\!-\!\Delta n)\}
\!\simeq\!\left[\exp\{\mathbf A\Delta\}\right]^{n}\!=\!\mathbf P_\Delta^{n},\!\!
\end{align}
where we have used the approximation $\exp\{\mathbf A(t-\Delta n)\}\simeq\mathbf I$, which holds  for $\Delta\ll 1$.
Moreover, since we assume $\Delta\ll 1$, we approximate the matrix exponential
$\mathbf P_\Delta=\exp\{\mathbf A\Delta\}$ with  the first order Taylor approximation
\begin{align}
\label{pdelta}
\mathbf P_\Delta\simeq\mathbf I+\Delta\mathbf A=
\mathbf I-\Delta\mathbf R(\mathbf I-\mathbf T)
\triangleq \tilde{\mathbf P}_\Delta.
\end{align}
Note that the approximation $\tilde{\mathbf P}_\Delta$ of $\mathbf P_\Delta$ is a feasible transition matrix with non-negative entries,
if $\Delta<\min_i\{1/R_i\}$. 
\subsection{Extension to $\mathbf s_E(t)$ piecewise constant}
\label{senonconstant}
In this section, we extend the previous analysis to the case where the external ambient state is piecewise constant,
\emph{i.e.}, $\mathbf s_E(t)=\mathbf s_{E,n},\ \forall n\in [\tau_n,\tau_{n+1}),\ \forall n\geq 0$, where $0=\tau_0$ and $\tau_n<\tau_{n+1},\ \forall n\geq 0$.
This analysis is of interest for the following experimental evaluation: the ED concentration is varied in order 
to measure the response in terms of fluctuations in the ATP level within the cell.

For this case, it is straightforward to derive the probability of the cell being in state $\mathbf s_I(t)=j\in\mathcal S_I\setminus\{\text{DEAD}\}$
at time $t\in [\tau_n,\tau_{n+1})$, for some $n\geq 0$, given $\mathbf s_I(0)=i\in\mathcal S_I\setminus\{\text{DEAD}\}$.
To this end, let $\mathbf T_n$ be the transition probability matrix within $\mathcal S_I\setminus\{\text{DEAD}\}$,
$\mathbf R_n$ be the rate matrix, and $\mathbf A_n=\mathbf R_n(\mathbf T_n-\mathbf I)$ be the flow matrix
 when $\mathbf s_E(t)=\mathbf s_{E,n}$.
Then, $\forall t\in [\tau_n,\tau_{n+1})$ we have
\begin{align*}
\mathbf P_t=\left[\prod_{m=0}^{n-1}\exp\{\mathbf A_m(\tau_{m+1}-\tau_m)\}\right]
\times \exp\{\mathbf A_n(t-\tau_n)\}.
\end{align*}
where we have defined $\prod_{m=0}^{n-1}\mathbf C_m=\mathbf C_0\times \mathbf C_1\times\cdots\times\mathbf C_{n-1}$,
and we have used the fact that, from the Markov property,
\begin{align*}
&\mathbb P(\mathbf s_I(t)=j|\mathbf s_I(0)=i)
=\!\!\!\!\!\!\!\!\!\!\!\!\!\!\!\!\!\!\!
\sum_{s_0,s_1,\dots,s_{n}\in\mathcal S_I\setminus\{\text{DEAD}\}}\!\!\!\!\!\!\!\!\!\!
\mathbb P(\mathbf s_I(t)=j|\mathbf s_I(\tau_n)=s_n)
\\&\times
\prod_{m=0}^{n-1}
\mathbb P(\mathbf s_I(\tau_{m+1})=s_{m+1}|\mathbf s_I(\tau_{m})=s_{m}),
\end{align*}
and, since $\mathbf s_E(\tau)$ is constant in the time interval $[\tau_m,\tau_{m+1})$,
the probability $\mathbb P(\mathbf s_I(\tau_{m+1})=s_{m+1}|\mathbf s_I(\tau_{m})=s_{m})$
can be computed as in Sec. \ref{TXprob}.
 \section{Application to cell-lifetime computation, Isolated cell}
 \label{celllifetime}
 For every cell in the bacterial chain, it is possible that, at some point in time, due to variations in the energetic state of the cell and changes to the supply of the ED and the EA, the cell reaches a state where its ATP consumption rate reaches a minimum value (\emph{e.g.}, zero). Once a cell enters this state, it is considered dead and its ATP consumption rate may not restore to normal values, thus jeopardizing the overall functionality of the cable. Accordingly, the time it takes for a cell to reach this irreversible state is defined as the \emph{lifetime of the cell}. This quantity can be measured experimentally by using indicators of cellular respiratory activity. In an experimental setup where cells in a bacterial chain can be characterized on an individual basis, cellular lifetime is one of the easiest measurable quantities that contains a significant amount of information regarding the specific properties of the target cell. In this section, we apply the stochastic model presented in Sec. \ref{isolatedcell} to the computation of the lifetime of an isolated cell, defined as follows.
 \begin{definition}
 The lifetime of the cell, $L$, is defined as
 \begin{align}
 L=\min\{t>0:\mathbf S_I(t)=\text{\em{DEAD}}\}.
 \end{align}
 \end{definition}
 Equivalently, letting $k^*=~\min\{k>0:\mathbf S_k=\text{DEAD}\}$, we have
 $L=T_{k^*}$.
 
 In this section, we compute the probability density function (pdf) of the lifetime, $f_L(t;\boldsymbol{\pi}_0)$, as well as the expected lifetime $\mathbb E[L|\boldsymbol{\pi}_0]$, given some initial state
 distribution $\boldsymbol{\pi}_0(i),i\in\mathcal S_I\setminus\{\text{DEAD}\}$.
$f_L(t;\boldsymbol{\pi}_0)$ is given by (we use $\mathbb P$ to denote also a~pdf)
 \begin{align}
&f_L(t;\boldsymbol{\pi}_0)
=\mathbb P(L=t|\boldsymbol{\pi}_0)
\\&
=\sum_{k=0}^{\infty}\mathbb P(L=t,\text{Death occurs at the $(k+1)$th event}|\boldsymbol{\pi}_0).
\nn
 \end{align}
 Note that the event $(L=t,$Death occurs at the $(k+1)$th event$)$ is equivalent to
 \begin{align}
 \mathbf S_k\in\mathcal S_I\setminus\{\text{DEAD}\},\mathbf S_{k+1}=\text{DEAD},T_{k+1}=t,
 \end{align}
 \emph{i.e.}, the cell is alive upon occurrence of the $k$th event, and dies upon occurrence of the $(k+1)$th event.
  Therefore, we obtain
 \begin{align}\label{lifetime}
f_L(t;\boldsymbol{\pi}_0)=
\sum_{k=0}^{\infty}\sum_{i\in\mathcal S_I\setminus\{\text{DEAD}\}}g_k(i,t),
 \end{align}
where we have defined $g_k(i,t)\triangleq\mathbb P(\mathbf S_k=i,\mathbf S_{k+1}=\text{DEAD},T_{k+1}=t|\boldsymbol{\pi}_0)$.
In order to compute $g_k(i,t)$, we first determine, for $k\geq 0$ and $s\in\mathcal S_I\setminus\{\text{DEAD}\}$,
 \begin{align}
&h_k(s,t)\triangleq \mathbb P(\mathbf S_k=s,T_k=t|\boldsymbol{\pi}_0).
 \end{align}
 For $k=0$, this is given by $h_0(s,t)=\boldsymbol{\pi}_0(s)\delta(t)$, where $\delta(t)$ is the Kronecker delta function.
For $k>0$, we have
 \begin{align}
&h_k(s,t)\!=\!
\sum_{j}\!\int_0^t \mathbb P(\mathbf S_k=s,T_k=t,\mathbf S_{k-1}\!=\!j,T_{k-1}\!=\tau|\boldsymbol{\pi}_0)\mathrm d\tau
\nonumber\\&
=
\sum_{j} \int_0^t \mathbb P(T_k-T_{k-1}=t-\tau|\mathbf S_{k-1}=j)
\nonumber\\&
\qquad\times
\mathbb P(\mathbf S_k=s|\mathbf S_{k-1}=j,T_{k-1}=\tau)
h_{k-1}(j,\tau)\mathrm d\tau
\nonumber\\&
=
\sum_{j} \int_0^t R_j e^{-R_j(t-\tau)}
\mathbf T(j,s)
h_{k-1}(j,\tau)\mathrm d\tau.
 \end{align}
It follows that
 \begin{align}
&g_k(s,t)=\mathbb P(\mathbf S_k=s,\mathbf S_{k+1}=\text{DEAD},T_{k+1}=t)
\nn\\&
=
\int_0^t  \mathbb P(\mathbf S_k=s,\mathbf S_{k+1}=\text{DEAD},T_{k+1}=t,T_k=\tau)\mathrm d\tau
\nonumber
\\&
=
\int_0^t  R_s e^{-R_s(t-\tau)}\left(1-\sum_j\mathbf T(s,j)\right)h_k(s,\tau)
\mathrm d\tau,
 \end{align}
 where $1-\sum_j\mathbf T(s,j)$ is the transition probability to state $\text{DEAD}$, from state $s$.
Substituting in (\ref{lifetime}), we obtain
 \begin{align*}
& f_L(t;\boldsymbol{\pi}_0)
\!=\!\sum_{s}\!\!\int_0^t  R_s e^{-R_s(t-\tau)}\!\!\left(1-\sum_j\mathbf T(s,j)\right)H(s,\tau)
\mathrm d\tau,
 \end{align*}
where we have defined 
 \begin{align}
&H(s,t)\triangleq\sum_{k=0}^{\infty}f_k(s,t)
\\&
=h_0(s,t)+
\sum_{j} \int_0^t R_j e^{-R_j(t-\tau)}\mathbf T(j,s)H(j,\tau)\mathrm d\tau
\nonumber\\&
=
\boldsymbol{\pi}_0(s)\delta(t)+
\sum_{j} \int_0^t R_j e^{-R_j(t-\tau)}\mathbf T(j,s)H(j,\tau)\mathrm d\tau.\nn
 \end{align}
Then, we obtain
 \begin{align*}
& \mathbb E[L]
\!=\!\int_0^{\infty}\!\!\!\!t\sum_{s}\int_0^t\!\!R_s e^{-R_s(t-\tau)}\!\!\left(\!\!1\!-\!\sum_j\mathbf T(s,j)\!\!\right)H(s,\tau)
\mathrm d\tau\mathrm dt
\nonumber\\&
\!=\!\sum_{s}\!\int_0^{\infty}\!\!\!
R_s e^{R_s\tau}\!\!\left(1-\sum_j\mathbf T(s,j)\!\!\right)\!\!H(s,\tau)
\int_\tau^\infty te^{-R_st}\mathrm d t
\mathrm d\tau.
 \end{align*}
Using the fact that $ \int_\tau^\infty te^{-R_st}\mathrm d t
 {=}
\frac{e^{-R_s\tau}}{R_s}\left(\tau{+}\frac{1}{R_s}\right)
$, we~obtain
 \begin{align}
& \mathbb E[L]
=
\sum_{s}\left(1-\sum_j\mathbf T(s,j)\right)\int_0^{\infty}\tau H(s,\tau)\mathrm d\tau
\nn\\&
\quad+\frac{1}{R_s}\sum_{s}\left(1-\sum_j\mathbf T(s,j)\right)\int_0^{\infty}H(s,\tau)\mathrm d\tau
\nonumber\\&
=
\sum_{s}\left(1-\sum_j\mathbf T(s,j)\right)Q(s)
\nn\\&
\quad+\sum_{s}\frac{1-\sum_j\mathbf T(s,j)}{R_s}\sum_{k=0}^\infty \mathbb P(S_k=s|\boldsymbol{\pi}_0),
 \end{align}
where we have defined $Q(s)\triangleq\int_0^{\infty}\tau H(s,\tau)\mathrm d\tau$.
This term can be computed as
\begin{align}
&Q(s)=\int_0^{\infty}t H(s,t)\mathrm dt
\nn\\&
=
\sum_{j} R_j\mathbf T(j,s)
\int_0^\infty e^{R_j\tau} H(j,\tau)
\int_\tau^\infty t e^{-R_jt}\mathrm dt
 \mathrm d\tau\nn
\\&=\nonumber
\sum_{j}\mathbf T(j,s)
\int_0^\infty H(j,\tau)\left(\tau+\frac{1}{R_j}\right)
 \mathrm d\tau
 \nn\\&
=
\sum_{j}\mathbf T(j,s)Q(j)
+
\sum_{j}\frac{\mathbf T(j,s)}{R_j}\sum_{k=0}^\infty \mathbb P(S_k=j|\boldsymbol{\pi}_0).
\end{align}
Let $\mathbf Q$ be a row vector with elements $Q(j)$, and $\mathbf x{=}(\mathbf I{-}\mathbf T)\mathbf 1$ be the column vector associated to transitions from 
the transient states to the DEAD state.
We obtain
\begin{align}
Q(s)=\mathbf Q\mathbf e_s
=
\mathbf Q\mathbf T\mathbf e_s
+
\boldsymbol{\pi}_0^T(\mathbf I-\mathbf T)^{-1}\mathbf R^{-1}\mathbf T\mathbf e_s,
\end{align}
where we have used the fact that
$
\sum_{k=0}^\infty \mathbb P(S_k=s|\boldsymbol{\pi}_0)
=
\boldsymbol{\pi}_0^T(\mathbf I-\mathbf T)^{-1}\mathbf e_s
$.
Therefore, we obtain
\begin{align}
\mathbf Q
=
\boldsymbol{\pi}_0^T(\mathbf I-\mathbf T)^{-1}\mathbf R^{-1}\mathbf T(\mathbf I-\mathbf T)^{-1}.
\end{align}
Substituting in the expression of the expected lifetime, we obtain
 \begin{align}
& \mathbb E[ L]
=
\boldsymbol{\pi}_0^T(\mathbf I-\mathbf T)^{-1}\mathbf R^{-1}(\mathbf I-\mathbf T)^{-1}\mathbf x.
 \end{align}
 Finally, we use the fact that $\mathbf x=(\mathbf I-\mathbf T)\mathbf 1$, yielding the expression of the expected lifetime
 \begin{align}
& \mathbb E[ L]
=
\boldsymbol{\pi}_0^T(\mathbf I-\mathbf T)^{-1}\mathbf R^{-1}\mathbf 1.
 \end{align}
 
\section{Parameter estimation and Experimental validation}
\label{estimation}
As an example of experiments related to our stochastic model, Ozalp et al. \cite{Ozalp} have measured \emph{in vivo} levels of ATP and NADH in the yeast \emph{Saccharomyces cerevisiae}, as they abruptly add ED to a suspension of starved yeast cells. Since we have theoretically investigated the single cell system, this work, which is performed on a culture of mutually-independent yeast cells, can be used as a test for our stochastic model.

Although the ETC in yeast is not exactly identical to the bacterial counterpart, the principles on which the model is based upon are conserved between yeast and bacteria. These include the involvement of an ED, electron carriers such as NADH, ETC and an EA, which in the case of yeast is molecular Oxygen. 

We have extracted the measured quantities from \cite{Ozalp} in the form of ATP and NADH concentrations as a function of time. In accordance with \cite{Ozalp}, we have assumed that yeast cells are initially starved and, at some point in time, the ED is added to the cell suspension. This triggers an increase in ATP and NADH production as well as ATP consumption. In extracting the data, we have averaged out the small oscillations in NADH and ATP concentrations in time, since these are mainly caused by an enzyme involved in the metabolic pathway that is specific to yeast and does not exist in the bacterial strain that we are interested in, \emph{Shewanella oneidensis} MR-1. Therefore, in matching the experimental data from yeast to our model, we have only taken into account the large scale variations of the levels of ATP and NADH over time.

Let $\{(\mathbf s_{I,k}^{i},t_k),\ k=0,1,\dots,N\}$ be  the time-series of the state of cell $i$
 at times $t_k$, where $0=t_0<t_1<\dots<t_N$.
Let $\mathbf s_E(t)$ be the known profile of the concentration of the external ED  and EA, which we assume to be piecewise constant,
as in Sec. \ref{senonconstant}, and the same for all cells.
In particular, we assume that $\mathbf s_E(t)=\mathbf s_{E,k},\ \forall t\in [t_k,t_{k+1}),\ \forall k=0,1,\dots, N-1$, so that the external state is constant in the time interval between two consecutive measurements.
The measurement collected in \cite{Ozalp} at time $t_k$ is
\begin{align}
\label{ts}
& \text{NADH}_k=\alpha_{NADH}\frac{1}{M}\sum_{i=1}^M m_{CH}^{i}(t_k)+\tilde{w}_k^{(NADH)},
\nonumber\\
& \text{ATP}_k=\alpha_{ATP}\frac{1}{M}\sum_{i=1}^M n_{ATP}^{i}(t_k)+\tilde{w}_k^{(ATP)},
\end{align}
where
$\text{NADH}_k$ is the measurement of NADH (typically, fluorescence level \cite{Ozalp}),
 whereas $\text{ATP}_k$ is the measurement of ATP (typically, in $\meas{mM}$ \cite{Ozalp});
 the constants $\alpha_{NADH}$ and $\alpha_{ATP}$ account for the conversion in the unit of measurements of NADH and ATP, respectively,
 from the stochastic model presented in this paper (electron units)
 to the experimental setup (fluorescence level and $\meas{mM}$, respectively);
 and $\tilde{w}_k^{(NADH)}$ and $\tilde{w}_k^{(ATP)}$
 are zero mean Gaussian noise samples, each i.i.d. over time, with variance
 $\sigma_{NADH}^2$ and $\sigma_{ATP}^2$, respectively. A practical assumption is that $M\gg~1$, so that 
$\frac{1}{M}\sum_{i=1}^M m_{CH}^{i}(t_k)\simeq\mathbb E[m_{CH}(t_k)]$
and $\frac{1}{M}\sum_{i=1}^M n_{ATP}^{i}(t_k)\simeq\mathbb E[n_{ATP}(t_k)]$,
where the expectation is computed with respect to the state distribution at time $t_k$, given by $\boldsymbol{\pi}_0^T\mathbf P_{t_k}$.  
 Letting
$\mathbf y_k=[\alpha_{NADH}^{-1}\text{NADH}_k,\alpha_{ATP}^{-1}\text{ATP}_k]$, $\mathbf w_k=[\alpha_{NADH}^{-1}\tilde{w}_k^{(NADH)},\alpha_{ATP}^{-1}\tilde{w}_k^{(ATP)}]$ and
 $\mathbf Z\in\mathbb  R^{(|\mathcal S_I|-1)\times 2}$ with $j$th row $[\mathbf Z]_{j,:}=[m_{CH}(j),n_{ATP}(j)]$, where $m_{CH}(j)$ and $n_{ATP}(j)$
are the NADH and ATP levels in the state corresponding to index $j$,
we thus obtain
\begin{align}
\label{obsmodel}
\mathbf y_k=\boldsymbol{\pi}_0^T\mathbf P_{t_k}\mathbf Z+\mathbf w_k\simeq
\boldsymbol{\pi}_0^T\prod_{j=1}^{k}\mathbf P_{\Delta,j-1}^{n_j-n_{j-1}}\mathbf Z+\mathbf w_k,
\end{align}
where $\prod_{j=1}^{k}\mathbf P_{\Delta,j-1}^{n_j-n_{j-1}}
=\mathbf P_{\Delta,0}^{n_1}\times
\mathbf P_{\Delta,1}^{n_2-n_1}
\times\dots\times
\mathbf P_{\Delta}^{n_k-n_{k-1}}$, $n_j=\lceil t_j/\Delta\rceil$, with $n_0=0$, and $\mathbf P_{\Delta,j-1}$ is the transition matrix with time-step size $\Delta$, when the external state
takes value $\mathbf s_{E,j-1}$.
In the last step, we have used the approximation (\ref{approx}).
Herein, we assume that $\mathbf w_k\sim\mathcal N(\mathbf 0,\sigma_w^2\mathbf I_2)$, \emph{i.e.},
$\alpha_{NADH}^{-2}\sigma_{NADH}^2=\alpha_{ATP}^{-2}\sigma_{ATP}^2=\sigma_w^2$.
\subsection{Parametric model}\label{parammodel}
The statistics of the system, defined by the transition probability matrix $\mathbf P_{t}$,
is determined by the rates $\lambda_{CH}$, $\mu_{CH}$ and $\mu_{ATP}$.
In this section, we present a parametric model for these rates, based on biological constraints.
Specifically, we let
\begin{align}\label{flows}
\!\!\!\!\!\!\begin{array}{l}
\lambda_{CH}(\mathbf s_I(t);\mathbf s_E(t))=\gamma\sigma_{D}(t)+\rho\left(1-\frac{m_{CH}(t)}{M_{CH}}\right)\sigma_{D}(t),\\
\mu_{CH}(\mathbf s_I(t);\mathbf s_E(t))=\zeta\left(1-\frac{n_{ATP}(t)}{N_{AXP}}\right),\\
\mu_{ATP}(\mathbf s_I(t);\mathbf s_E(t))=\beta\sigma_{D}(t),
\end{array}\!\!
\end{align}
 where  $\gamma,\rho,\zeta,\beta\in\mathbb R_+$ are parameters, that we want  to estimate,
 and $\mathbb R_+$ is the set of non-negative reals.
The NADH generation rate $\lambda_{CH}$ primarily depends on the concentration of available ED, as explained in Sec. \ref{constr}.
 Additionally, it depends on the number of available NAD molecules in the cell, since the ED reacts with NAD to form NADH. The more NAD molecules are available, the higher the rate of the NADH-producing reaction. 
Moreover, the larger the ATP level,
the smaller the ATP generation rate $\mu_{CH}(\mathbf s_I(t);\mathbf s_E(t))$. This is true because the ATP synthase, the protein responsible for ATP production, transforms ADP into ATP. Since the sum of ATP and ADP molecules in the cell is conserved, a higher ATP level corresponds to a lower ADP level. Therefore, as there is more ATP available in the cell, there are less ADP molecules available for ATP synthase to produce additional ATP molecules, which, in turn, results in a smaller ATP production rate.
Finally, the larger the ED concentration, the larger the ATP consumption rate $\mu_{ATP}(\mathbf s_I(t);\mathbf s_E(t))$. This is shown to be true experimentally, for instance in \cite{Ozalp}. The reason behind this correlation is that cellular operations that consume ATP (\emph{e.g.}, ATP-ases) are directly regulated by the ED concentration. 
 Note that $\lambda_{CH}$, $\mu_{CH}$ and $\mu_{ATP}$ further need to satisfy the constraints listed in Sec. \ref{constr}.
We assume that all the cells are alive throughout the experiment, and set the death rate $\delta(\mathbf s_I(t);\mathbf s_E(t))=0$.

We define the parameter vector $\mathbf x=[\gamma,\rho,\zeta,\beta]$, which is estimated via maximum likelihood (ML) in the next section.
Therefore, the flow matrix $\mathbf A$, defined in (\ref{p4}),
is a linear function of the entries of $\mathbf x$. We write such a dependence as
$\mathbf A(\mathbf x,\mathbf s_{E})$.
Similarly, from (\ref{pdelta}), we write
$\mathbf P_\Delta(\mathbf x,\mathbf s_E)=\mathbf I+\Delta\mathbf A(\mathbf x,\mathbf s_E)$.
\subsection{Maximum Likelihood estimate of $\mathbf x$}
For a given time series $\{(\mathbf y_k,t_k),\ k=0,1,\dots,N\}$, and the piecewise constant profile of the external state $\mathbf s_E(t)$,
 in this section we design a ML estimator of $\mathbf x$.
 Since the initial distribution $\boldsymbol{\pi}_0$ is unknown, we also estimate it jointly with $\mathbf x$.
 Note that, since the death rate is zero, the entries of $\boldsymbol{\pi}_0$ need to sum to one, \emph{i.e.}, $\mathbf 1^T\boldsymbol{\pi}_0=1$.
 Moreover, we further enforce the constraints $\boldsymbol{\pi}_0^T\mathbf Z=\mathbf y_0$, \emph{i.e.}, the expected values of the NADH and ATP pools
 at time $t_0$ equal the measurement $\mathbf y_0$. Therefore, we have the linear equality constraint $\boldsymbol{\pi}_0^T[\mathbf Z,\mathbf 1]=[\mathbf y_0,1]$,
 and the inequality constraint $\boldsymbol{\pi}_0\geq 0$ (component-wise). We denote the constraint set as $\mathcal P$, so that $\boldsymbol{\pi}_0\in\mathcal P$.
    Due to the Gaussian observation model (\ref{obsmodel}),
 the ML estimate of $(\mathbf x,\boldsymbol{\pi}_0)$ is given by
 \begin{align}
 \label{ML}
(\hat{\mathbf x},\hat{\boldsymbol{\pi}}_0)=\underset{\mathbf x\geq 0,\boldsymbol{\pi}_0\in\mathcal P}{\arg\min} f(\mathbf x,\boldsymbol{\pi}_0),
 \end{align}
 where we have defined the negative log-likelihood cost function
\begin{align*}
f(\mathbf x,\boldsymbol{\pi}_0)\triangleq
\frac{1}{2}\sum_{k=0}^N \left\|\mathbf y_k-
\boldsymbol{\pi}_0^T
\prod_{j=1}^{k}\mathbf P_{\Delta}(\mathbf x,\mathbf s_{E,j-1})^{n_j-n_{j-1}}
\mathbf Z\right\|_F^2\!\!.
\end{align*}
For a fixed $\mathbf x$, the optimization over $\boldsymbol{\pi}_0$ is a quadratic programming problem, which can be solved efficiently using, \emph{e.g.}, interior-point methods \cite{Boyd,Boggs}. On the other hand, for fixed $\boldsymbol{\pi}_0$, the optimization over $\mathbf x$ is a non-convex optimization problem.
Therefore, we resort to a gradient descent (GD) algorithm to optimize over $\mathbf x$, which only guarantees convergence to a \emph{local} optimum.
Finally, we employ an iterative method to solve (\ref{ML}), \emph{i.e.}, we optimize over 
$\boldsymbol{\pi}_0$ for the current estimate of $\mathbf x$, then we optimize over $\mathbf x$ for the current estimate of $\boldsymbol{\pi}_0$, and so on.
The derivative of $f(\mathbf x,\boldsymbol{\pi}_0)$ with respect to $\mathbf x_j$ is given~by
\begin{align}
&
\left[\nabla_{\mathbf x}f(\mathbf x,\boldsymbol{\pi}_0)\right]_j=
\frac{\mathrm d}{\mathrm d\mathbf x_j}f(\mathbf x,\boldsymbol{\pi}_0)
\nonumber\\&
=
-\sum_{k=0}^N\boldsymbol{\pi}_0^T\frac{\mathrm d
\left[\prod_{j=1}^{k}\mathbf P_{\Delta}(\mathbf x,\mathbf s_{E,j-1})^{n_j-n_{j-1}}\right]
}{\mathrm d\mathbf x_j}\mathbf Z
\nn\\&
\times
\left(\mathbf y_k-\boldsymbol{\pi}_0^T\prod_{j=1}^{k}\mathbf P_{\Delta}(\mathbf x,\mathbf s_{E,j-1})^{n_j-n_{j-1}}
\mathbf Z\right)^T.
\end{align}
We further assume that the intervals satisfy $t_{k+1}-t_k=T,\ \forall k$, so that $n_k=kn,\ \forall k$, and we enforce $n=2^b$, for some integer $b>0$.
This can be accomplished by appropriately choosing $\Delta\ll 1$.
Then, the derivative
$\frac{\mathrm d
\left[\prod_{j=1}^{k}\mathbf P_{\Delta}(\mathbf x,\mathbf s_{E,j-1})^{n}\right]
}{\mathrm d\mathbf x_j}$
can be efficiently computed recursively as
\begin{align*}
&\frac{\!\mathrm d\!\!\left[\overset{k}{\underset{j=1}{\prod}}\!\mathbf P_{\!\Delta\!}(\mathbf x,\mathbf s_{E,j-1})^{n}\right]\!}{\mathrm d\mathbf x_j}
{=}
\frac{\!\mathrm d\!\!\left[\overset{k-1}{\underset{j=1}{\prod}}\mathbf P_{\!\Delta\!}(\mathbf x,\mathbf s_{E,j-1})^{n}\right]\!}{\mathrm d\mathbf x_j}\mathbf P_{\!\Delta\!}(\mathbf x,\mathbf s_{E,k-1})^{n}
\nonumber\\&
+
\prod_{j=1}^{k-1}\mathbf P_{\Delta}(\mathbf x,\mathbf s_{E,j-1})^{n}\frac{\mathrm d\mathbf P_{\Delta}(\mathbf x,\mathbf s_{E,k-1})^{n}}{\mathrm d\mathbf x_j},
\end{align*}
where the derivative $\frac{\mathrm d\mathbf P_{\Delta}(\mathbf x,\mathbf s_{E})^{n}}{\mathrm d\mathbf x_j}$ can 
be efficiently computed recursively as
\begin{align}
&\frac{\mathrm d\mathbf P_{\Delta}(\mathbf x,\mathbf s_{E})^{2^{b}}}{\mathrm d\mathbf x_j}
=
\frac{\mathrm d\mathbf P_{\Delta}(\mathbf x,\mathbf s_{E})^{2^{b-1}}}{\mathrm d\mathbf x_j}\mathbf P_{\Delta}(\mathbf x,\mathbf s_{E})^{2^{b-1}}
\nn\\&\quad
+
\mathbf P_{\Delta}(\mathbf x,\mathbf s_{E})^{2^{b-1}}
\frac{\mathrm d\mathbf P_{\Delta}(\mathbf x,\mathbf s_{E})^{2^{b-1}}}{\mathrm d\mathbf x_j}.
\end{align}
Finally,
let $\hat{\mathbf x}_p$ be the estimate of $\mathbf x$ at the $p$th iteration of the GD algorithm.
Then, the GD algorithm updates the ML estimate of $\mathbf x$ as
\begin{align}
&\hat{\mathbf x}_{p+1}=(\hat{\mathbf x}_p-\mu_p\nabla_{\mathbf x} f(\hat{\mathbf x}_p,\hat{\mathbf e}_p))^+,
\end{align}
where $0<\mu_k\ll 1$ is the (possibly, time-varying) step size
and we have defined $(v)^+=\max\{v,0\}$, applied to each entry, so that 
a non-negativity constraint is enforced (in fact, the entries of $\mathbf x$ need to be non-negative).
\subsection{Results}
\begin{figure*}
\begin{center}
\setlength{\tabcolsep}{1mm}
\begin{tabular}{cc}
\subfigure[
Prediction of expected ATP level over time (in $\meas{mM}$) and experimental time-series.
]{\includegraphics*[width=0.48\linewidth]{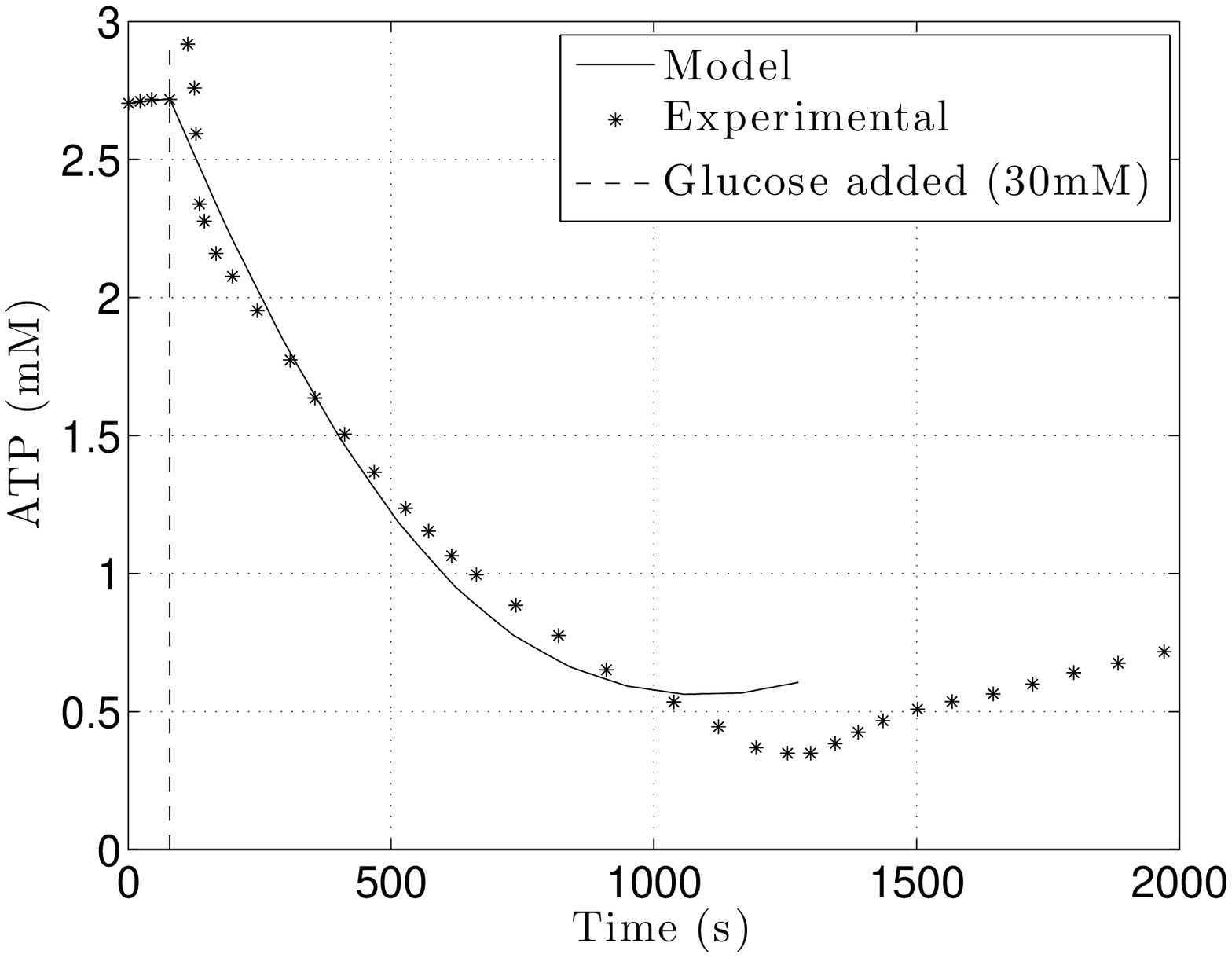}}&
\subfigure[
Prediction of expected NADH level over time (equivalent fluorescence level) and experimental time-series.
]{\includegraphics*[width=0.48\linewidth]{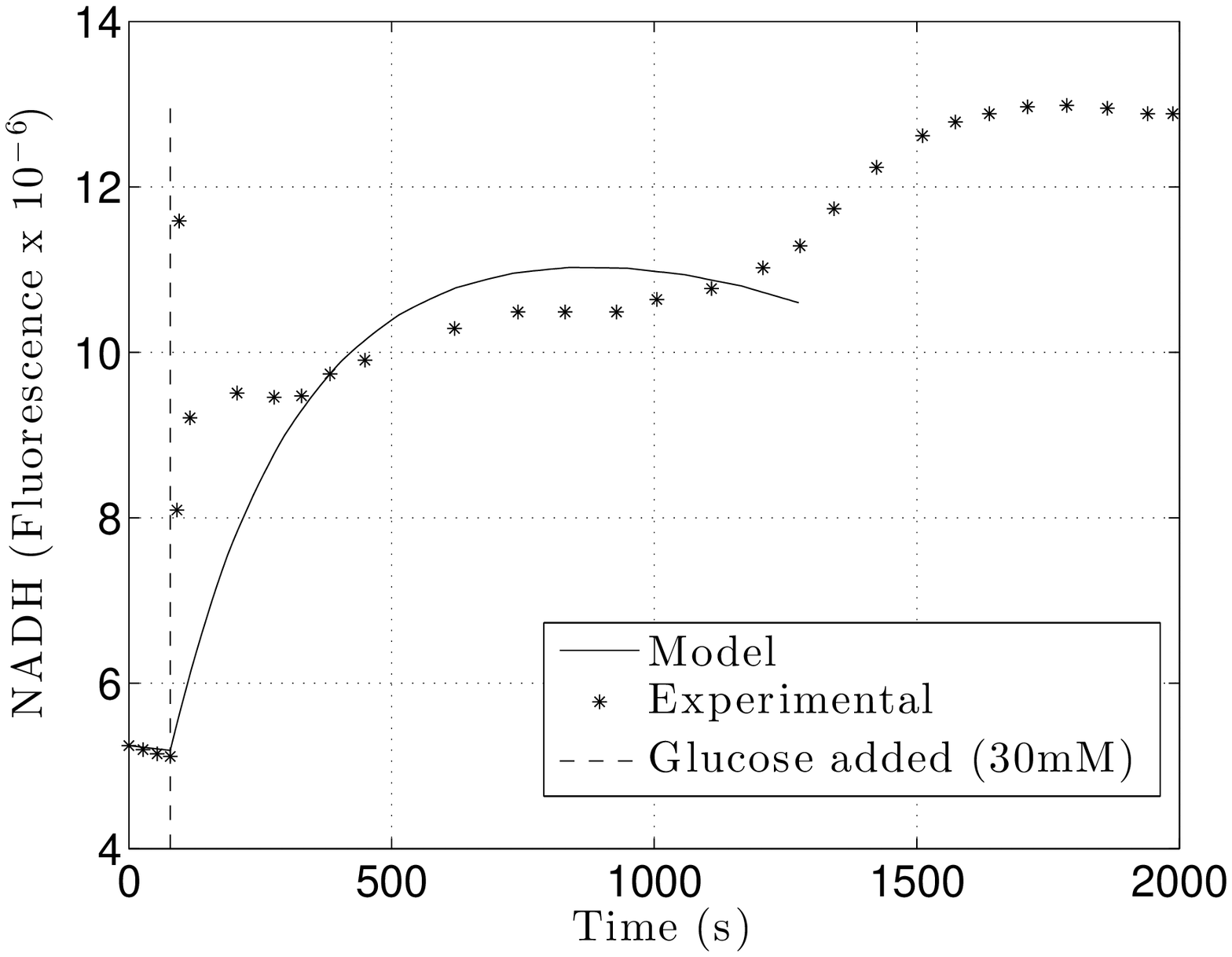}}
\\
\subfigure[
Prediction of expected ATP synthesis and consumption rates per cell.
]{\includegraphics*[width=0.48\linewidth]{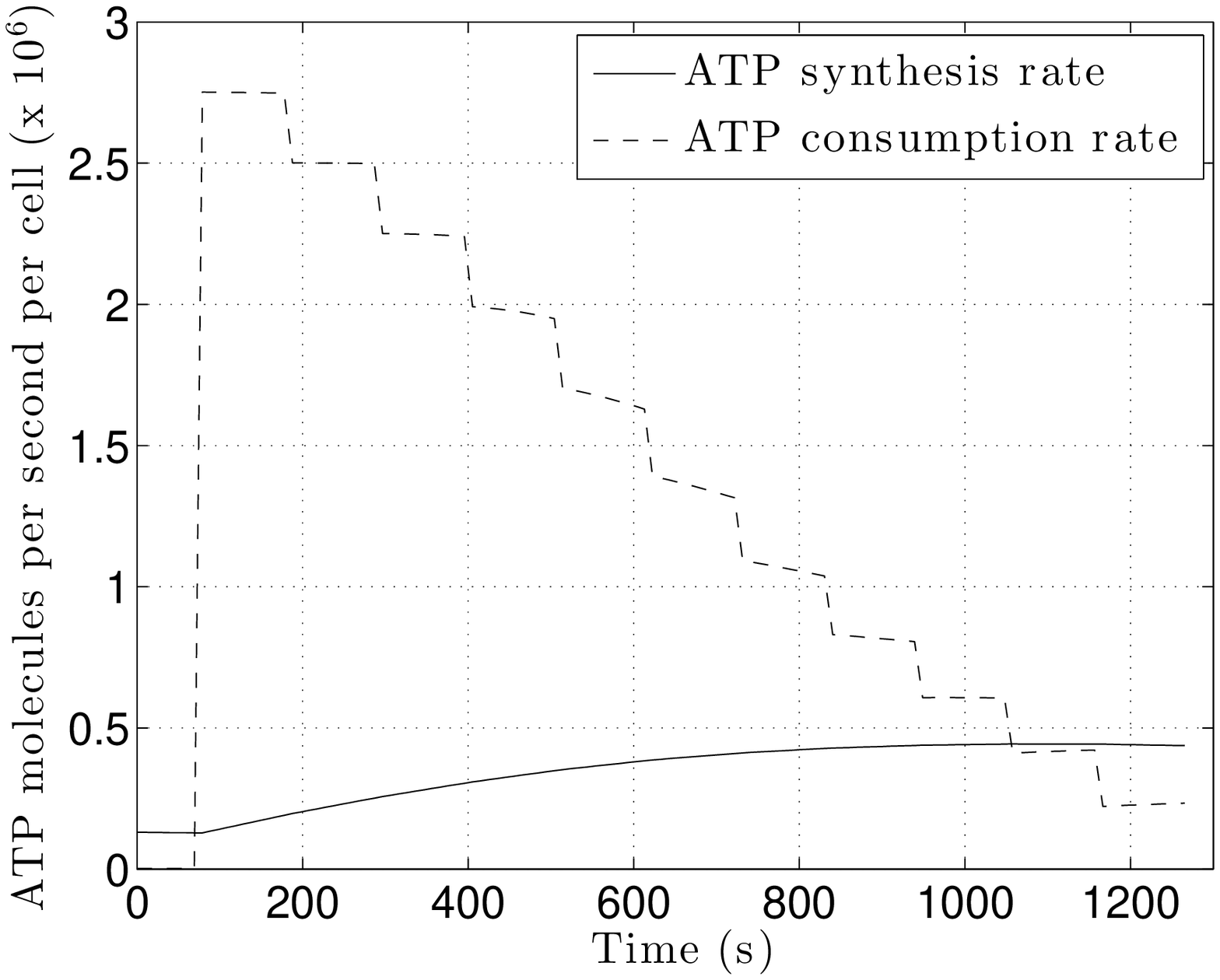}}&
\subfigure[
Prediction of expected NADH generation and consumption rates per cell.
]{\includegraphics*[width=0.48\linewidth]{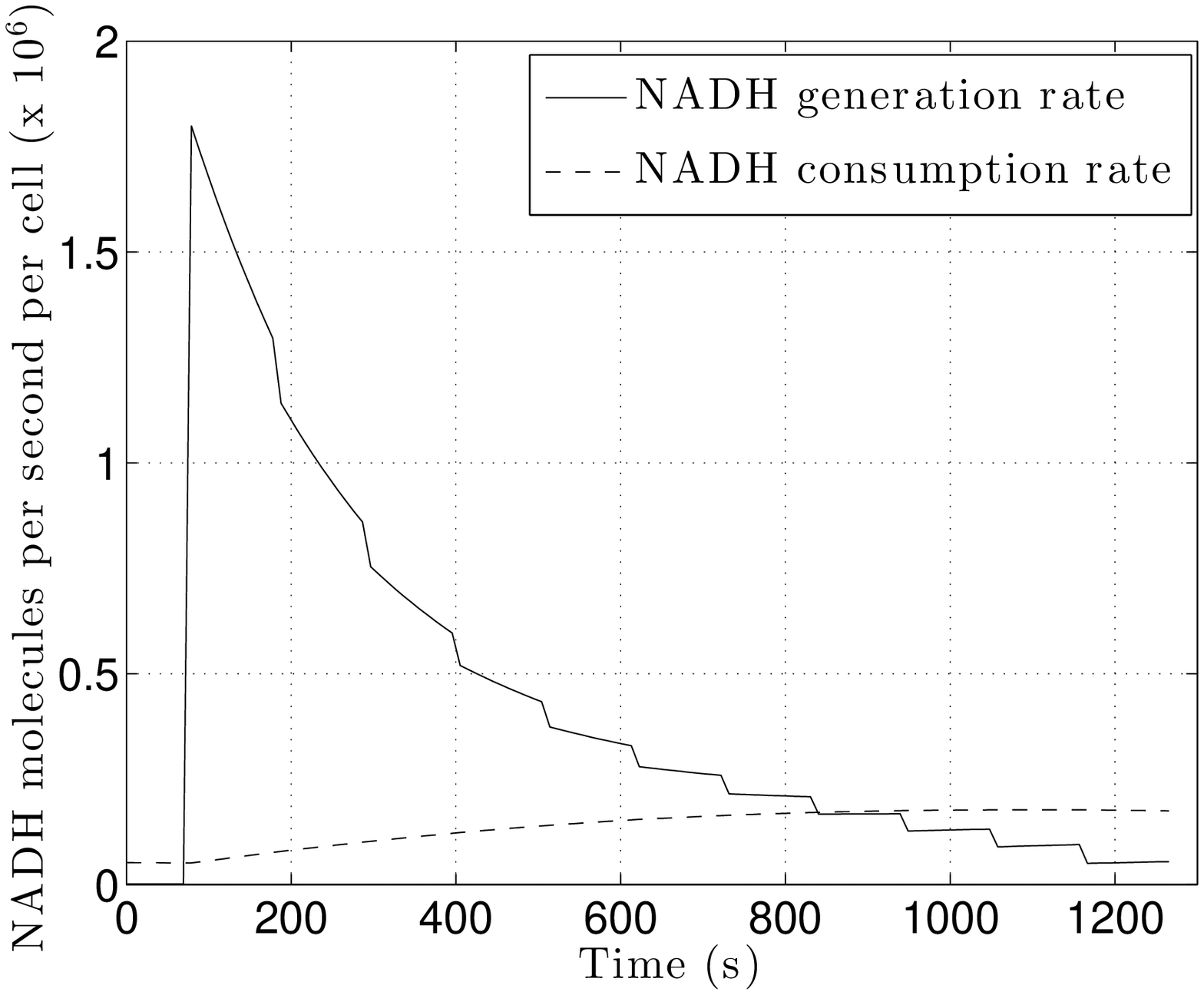}}
\end{tabular}
\caption{\vspace{-4mm}}
\label{figmodel}
\end{center}
\vspace{-7mm}
\end{figure*}

We use the algorithm outlined above to fit the parameter vector $\mathbf x$ to the experimental data.
While, in principle, the capacities of both the IECP ($M_{CH}$) and the ATP pool ($N_{AXP}$) need to be estimated,
we found that $M_{CH}=N_{AXP}=20$ provides a good fit, and good trade-off between convergence of the estimation algorithm and fitting.
Note that the ATP capacity of the cell culture is $3.6\meas{mM}$ and
the concentration of cells is $10^{12} \meas{cells/liter}$ (see \cite{Ozalp}).
It follows that the capacity of the ATP pool of each cell is $2.16 \times10^{9}\meas{molecules/cell}$. Since, 
 approximately, $2.5$ ATP molecules are created by the flow of $2$ electrons in the ETC (see \cite[Sec. 18.6]{Biochemistry}),
 the ATP pool may carry 
$1.728 \times10^{9}\meas{electrons/cell}$. Therefore, one "unit" in the stochastic model corresponds to $N_E=0.864\times10^{8}\meas{electrons}$,
 or $1.08\times10^{8}\meas{ATP\ molecules}$.
 Similarly, since each NADH molecule carries $2$ electrons which actively participate in the ETC, we have that one "unit" corresponds to 
 $0.432\times10^{8}\meas{NADH\ molecules}$.
The time-series $\{(\text{ATP}_k,\text{NADH}_k,t_k)\}$ is first extracted from \cite{Ozalp},
where $\text{ATP}_k$ is in $\meas{mM}$ (stars in Fig. \ref{figmodel}.a), $\text{NADH}_k$ is a fluorescence level ($\times 10^{-6}$, stars in Fig. \ref{figmodel}.a), which we assume to be linearly proportional to the NADH level.
The time-series is then converted to feasible values in the stochastic model.
In particular, since the ATP capacity of the cell culture is $3.6\meas{mM}$, and assuming that 
the IECP capacity of the cell culture is $\max_k \text{NADH}_k=12.985$ [fluorescence $\times 10^{-6}$],
\emph{i.e.}, the maximum level reached in the NADH measurements, the time-series is
\begin{align}\label{conversion}
&y_k^{(NADH)}=\frac{\text{NADH}_k\ \meas{[fluorescence\ x\ 10^{-6}]}}{12.985\ \meas{[fluorescence\ x\ 10^{-6}]}}M_{CH},
\nn\\&
y_k^{(ATP)}=\frac{\text{ATP}_k\ \meas{[mM]}}{3.6\ \meas{[mM]}}N_{AXP},
\end{align}
so that, from (\ref{ts}), $\alpha_{NADH}\simeq 0.650\meas{[fluorescence\ x\ 10^{-6}]}$ and $\alpha_{ATP}=0.18\meas{[mM]}$.
Note that both $y_k^{(NADH)}$ and $y_k^{(ATP)}$ are dimensionless quantities.
The time-series $\{(\mathbf y_k,t_k)\}$, where $\mathbf y_k=[y_k^{(NADH)},y_k^{(ATP)}]$, is then fed into the estimation algorithm.
The EA concentration (molecular Oxygen) is assumed to be constant throughout the experiment, and sufficient to sustain reduction.
On the other hand, the ED concentration profile, extrapolated from \cite{Ozalp}, is zero
at time $t<80\meas{s}$, when cells are starved, $30\meas{mM}$ at $t=80\meas{s}$, when glucose is added to the starved cells,
and constantly decreases until it becomes zero at time $t\simeq 1300\meas{s}$, when cells become starved again.
\begin{remark}\label{rem1}
In the parameter estimation, the samples after $t\simeq 1300\meas{s}$ are discarded, since
cells are starved after that time, which, in turn, results in increased cell lysis occurring in the cell suspension. The cellular material released by lysis can be used by other intact cells as ED or EA. For this reason, and since the extent of cell lysis in unpredictable in the cell culture, the concentrations of the ED and the EA cannot be accurately determined after $t\simeq 1300\meas{s}$, rendering the corresponding experimental data useless.
\end{remark}

With this approach, the estimated parameters, used in (\ref{flows}) to compute the corresponding flow rates, are given by
\begin{align}
\left\{
\begin{array}{ll}
\hat\gamma=0&\meas{units/mM/s},\\
 \hat\rho=2.31\times 10^{-3}& \meas{units/mM/s},\\
\hat\zeta=4.866\times 10^{-3}&\meas{units/s},\\
\hat\beta=0.850\times 10^{-3}&\meas{units/mM/s},
\end{array}
\right.
\end{align}
where "units", equivalent to $0.864\times10^{8}\meas{electrons}$, refers to the number of slots being occupied/emptied in the respective queue of the  stochastic model,
and $\meas{mM}$ refers to the glucose ED concentration.
In particular, the "units" can be converted back into the original scale (\ref{conversion}), which is related to the overall cell culture.
The corresponding quantity related to a single cell is then obtained by
converting one "unit" to the corresponding molecular quantity (1 unit=$1.08\times10^{8}\meas{ATP\ molecules}$=$0.432\times10^{8}\meas{NADH\ molecules}$).

Figs. \ref{figmodel}.a and \ref{figmodel}.b plot, respectively, the ATP and NADH time-series, related to the cell culture, and the predicted values based on our proposed stochastic model.
We observe a good fit in the time-interval $t\in [0\meas{s},1300\meas{s}]$.
{The corresponding standard deviation of the error between the prediction and experimental curves is 0.1325[mM] and 1.7736[Fluorescence$\times 10^{-6}$], respectively.
The prediction error observed in the two figures can be explained by both 
cell lysis occurring in the bacterial population and the resulting distortion in the ATP and NADH levels,
 as explained in Remark \ref{rem1},
 and by the \emph{bias} introduced by our specific choice of the parametric model (\ref{flows}), which may not be sufficiently accurate to capture higher order fluctuations.
 The investigation of other parametric models to improve the prediction accuracy is left for future work.}
 Figs. \ref{figmodel}.c and \ref{figmodel}.d plot, respectively, the expected ATP and NADH generation and consumption rates over time, related to a single cell.
These biophysical parameters were not directly measured in the experiments by \cite{Ozalp}, 
but can be predicted by our proposed model.
We notice that the ATP generation/consumption rate is of the order of $5\times 10^5$/$3\times 10^6$, whereas the NADH generation/consumption rate
is of the order of $2\times 10^6$/$2\times 10^5$ (molecules per cell per second).
These values are indeed physical, and consistent with known metabolic rates in yeast \cite{Teusink}, which further motivates the development of this stochastic model as a predictive tool for microbial energetics.
\section{Future work}
\label{future}
Our current experimental work involves measuring ATP and NADH levels in isolated single cells of bacterium \emph{S. oneidensis} MR-1.
This organism is capable of extracellular electron transfer by utilizing its array of outer-membrane multi-heme cytochromes and, due to its unique properties, presents a great model organism for this study.
Similarly to the previous experiment done on the yeast \emph{Saccharomyces cerevisiae} \cite{Ozalp}, whose data have been used in our experimental validation in Sec. \ref{estimation}, the ED will be abruptly added to a culture of starved bacterial cells, 
 and, subsequently, the cellular ATP and NADH levels will be measured over time.
  As opposed to \cite{Ozalp}, we are
   developing an experimental setup in which the amount of the ED and the EA can be maintained at any desired level, therefore producing additional unprecedented data to be compared with our theoretical model.
      
{The next step in our experiments will be to assemble bacterial cables,
\emph{e.g.}, using similar techniques as in \cite{Wang},
 and perform ATP and NADH measurements in these cells as the availability and the type of the ED and the EA is varied. 
In order to control and stimulate the growth of bacterial cables,
 a population of cells will be initially grown in an ED/EA rich medium, and subsequently moved to an environment with limited amounts of ED/EA, causing
 cell growth to stop. Then, the cells will be placed in a microfluidic medium where they can be moved, \emph{e.g.}, via optical tweezers \cite{Liu2}, to form a one-dimensional bacterial cable.
 Adjusting the environmental parameters will then induce the production of electron transfer components in the bacteria, thus enabling long-range electron transfer in the cable. 
By keeping the concentration of ED/EA along the cable small, the bacteria are forced to use the externally provided solid-state electrodes as the electron source/sink and maintain the collective electron transfer through the cable.  The role of each cell in this collective behavior is in the form of establishing direct cell-cell contact and facilitating electron transfer to and from the adjacent cells, and cooperation of every single cell in the system is necessary to provide enough ED/EA to sustain the entire network.
 Solid-phase electrodes poised to a desired electric potential can be used as the EA for such a cable. The rate of electron transfer to such an electrode can be controlled by adjusting its potential, and this electron transfer rate to the electrode can be accurately measured. Similar manipulations of the ED and the EA and their availability and the subsequent measurements on the state of the cells and the transfer rates within the chain will result in a vast amount of quantitative data to validate our stochastic model for IET.}
\section{Conclusions}
\label{conclu}
In this paper, we have presented a stochastic model for electron transfer in bacterial cables.
 In particular,
we have specialized the stochastic model to the case of an isolated cell,
which is the building block of more complex bacterial cables,
 and we have provided an example of the application to the computation of the cell's lifetime.
Moreover, we have designed a parameter estimation framework, based on a parametric description of the model, guided by biological constraints.
The parameters were fit to experimental data available in the literature, demonstrating the capability of the proposed stochastic model to predict
salient features related to the energetic state of the cells, such as ATP generation and consumption rates.
This study is a first step towards addressing questions of more communications theoretic relevance, such as the interplay between information capacity of a microbial community and lifetime of the cells, reliability and delay in electron-based nanonetworks.

\bibliographystyle{IEEEtran}
\bibliography{IEEEabrv,Refs} 

\begin{thebibliography}{10}
\providecommand{\url}[1]{#1}
\csname url@samestyle\endcsname
\providecommand{\newblock}{\relax}
\providecommand{\bibinfo}[2]{#2}
\providecommand{\BIBentrySTDinterwordspacing}{\spaceskip=0pt\relax}
\providecommand{\BIBentryALTinterwordstretchfactor}{4}
\providecommand{\BIBentryALTinterwordspacing}{\spaceskip=\fontdimen2\font plus
\BIBentryALTinterwordstretchfactor\fontdimen3\font minus
  \fontdimen4\font\relax}
\providecommand{\BIBforeignlanguage}[2]{{%
\expandafter\ifx\csname l@#1\endcsname\relax
\typeout{** WARNING: IEEEtran.bst: No hyphenation pattern has been}%
\typeout{** loaded for the language `#1'. Using the pattern for}%
\typeout{** the default language instead.}%
\else
\language=\csname l@#1\endcsname
\fi
#2}}
\providecommand{\BIBdecl}{\relax}
\BIBdecl

\bibitem{CISS}
N.~Michelusi, S.~Pirbadian, M.~Y. El-Naggar, and U.~Mitra, ``{A model for
  electron transfer and cell energetics in bacterial cables},'' in \emph{48th
  Annual Conference on Information Sciences and Systems (CISS)}, March 2014,
  pp. 1--6.

\bibitem{Bassler}
M.~B. Miller and B.~L. Bassler, ``Quorum sensing in bacteria,'' \emph{Annual
  Review of Microbiology}, vol.~55, no.~1, pp. 165--199, 2001.

\bibitem{Visick15082005}
\BIBentryALTinterwordspacing
K.~L. Visick and C.~Fuqua, ``{Decoding Microbial Chatter: Cell-Cell
  Communication in Bacteria},'' \emph{Journal of Bacteriology}, vol. 187,
  no.~16, pp. 5507--5519, 2005. [Online]. Available:
  \url{http://jb.asm.org/content/187/16/5507.short}
\BIBentrySTDinterwordspacing

\bibitem{Nealson}
K.~H. K.~H.~Nealson, T.~Platt, and J.~W. Hastings, ``{Cellular Control of the
  Synthesis and Activity of the Bacterial Luminescent System},'' \emph{J.
  Bacteriol.}, vol. 104, no.~1, pp. 313--322, Oct. 1970.

\bibitem{P1906}
\BIBentryALTinterwordspacing
{IEEE P1906.1 -- Recommended Practice for Nanoscale and Molecular Communication
  Framework}. [Online]. Available:
  \url{https://standards.ieee.org/develop/project/1906.1.html}
\BIBentrySTDinterwordspacing

\bibitem{bush2010nanoscale}
S.~F. Bush, \emph{{Nanoscale Communication Networks}}.\hskip 1em plus 0.5em
  minus 0.4em\relax Artech House, 2010.

\bibitem{Nakano}
T.~Nakano, A.~W. Eckford, and T.~Haraguchi, \emph{{Molecular
  Communication}}.\hskip 1em plus 0.5em minus 0.4em\relax Cambridge University
  Press, 2013.

\bibitem{Akyildiz1}
\BIBentryALTinterwordspacing
I.~F. Akyildiz, F.~Brunetti, and C.~Bl\'{a}zquez, ``{Nanonetworks: A new
  communication paradigm},'' \emph{Computer Networks}, vol.~52, no.~12, pp.
  2260--2279, Aug. 2008. [Online]. Available:
  \url{http://dx.doi.org/10.1016/j.comnet.2008.04.001}
\BIBentrySTDinterwordspacing

\bibitem{Rose}
I.~S. Mian and C.~Rose, ``Communication theory and multicellular biology,''
  \emph{Integr. Biol.}, vol.~3, pp. 350--367, 2011.

\bibitem{Eckford07}
A.~Eckford, ``{Nanoscale Communication with Brownian Motion},'' in \emph{41st
  Annual Conference on Information Sciences and Systems (CISS)}, 2007, pp.
  160--165.

\bibitem{Kadloor}
S.~Kadloor, R.~Adve, and A.~Eckford, ``{Molecular Communication Using Brownian
  Motion With Drift},'' \emph{IEEE Transactions on NanoBioscience}, vol.~11,
  no.~2, pp. 89--99, 2012.

\bibitem{Einolghozati}
A.~Einolghozati, M.~Sardari, A.~Beirami, and F.~Fekri, ``{Capacity of discrete
  molecular diffusion channels},'' in \emph{IEEE International Symposium on
  Information Theory Proceedings (ISIT)}, 2011, pp. 723--727.

\bibitem{Fekri}
------, ``{Data gathering in networks of bacteria colonies: Collective sensing
  and relaying using molecular communication},'' in \emph{INFOCOM Workshops},
  2012, pp. 256--261.

\bibitem{Arjmandi}
H.~Arjmandi, A.~Gohari, M.~Kenari, and F.~Bateni, ``{Diffusion-Based
  Nanonetworking: A New Modulation Technique and Performance Analysis},''
  \emph{IEEE Communications Letters}, vol.~17, no.~4, pp. 645--648, 2013.

\bibitem{Mosayebi}
R.~Mosayebi, H.~Arjmandi, A.~Gohari, M.~N. Kenari, and U.~Mitra, ``{On bounded
  memory decoder for molecular communications},'' in \emph{Information Theory
  and Applications Workshop (ITA)}, Feb. 2014.

\bibitem{Oiwa}
M.~Moore, T.~Suda, and K.~Oiwa, ``{Molecular Communication: Modeling Noise
  Effects on Information Rate},'' \emph{IEEE Transactions on NanoBioscience},
  vol.~8, no.~2, pp. 169--180, 2009.

\bibitem{Kuran201086}
M.~{\c S}. Kuran, H.~B. Yilmaz, T.~Tugcu, and B.~{\"O}zerman, ``{Energy model
  for communication via diffusion in nanonetworks},'' \emph{Nano Communication
  Networks}, vol.~1, no.~2, pp. 86--95, 2010.

\bibitem{Reguera2}
G.~Reguera, ``{When microbial conversations get physical},'' \emph{Trends in
  Microbiology}, vol. 19(3), pp. 105--113, 2011.

\bibitem{Pfeffer}
C.~Pfeffer \emph{et~al.}, ``{Filamentous bacteria transport electrons over
  centimetre distances},'' \emph{Nature}, vol. 491(7423), pp. 218--221, 2012.

\bibitem{Kato}
S.~Kato, K.~Hashimoto, and K.~Watanabe, ``{Microbial interspecies electron
  transfer via electric currents through conductive minerals},''
  \emph{Proceedings of the National Academy of Sciences}, vol. 109, no.~25, pp.
  10\,042--10\,046, June 2012.

\bibitem{Gunduz}
D.~Gunduz, K.~Stamatiou, N.~Michelusi, and M.~Zorzi, ``{Designing intelligent
  energy harvesting communication systems},'' \emph{IEEE Communications
  Magazine}, vol.~52, no.~1, pp. 210--216, Jan. 2014.

\bibitem{MichelusiEH}
N.~Michelusi, K.~Stamatiou, and M.~Zorzi, ``{Transmission Policies for Energy
  Harvesting Sensors with Time-Correlated Energy Supply},'' \emph{IEEE
  Transactions on Communications}, vol.~61, no.~7, pp. 2988--3001, 2013.

\bibitem{Liu}
L.~Liu, R.~Zhang, and K.-C. Chua, ``{Wireless Information Transfer with
  Opportunistic Energy Harvesting},'' \emph{IEEE Transactions on Wireless
  Communications}, vol.~12, no.~1, pp. 288--300, 2013.

\bibitem{Naggar}
M.~Y. El-Naggar and S.~E. Finkel, ``{Live Wires: Electrical Signaling Between
  Bacteria},'' \emph{The Scientist}, vol.~27, no.~5, pp. 38--43, 2013.

\bibitem{Naggar2}
M.~Y. El-Naggar, G.~Wanger, K.~M. Leung, T.~D. Yuzvinsky, G.~Southam, J.~Yang,
  W.~M. Lau, K.~H. Nealson, and Y.~A. Gorby, ``{Electrical transport along
  bacterial nanowires from Shewanella oneidensis MR-1},'' \emph{Proceedings of
  the National Academy of Sciences}, vol. 107(42), pp. 18\,127--18\,131, 2010.

\bibitem{Pirbadian}
S.~Pirbadian and M.~Y. El-Naggar, ``{Multistep hopping and extracellular charge
  transfer in microbial redox chains},'' \emph{Physical Chemistry Chemical
  Physics}, vol.~14, pp. 13\,802--13\,808, 2012.

\bibitem{Reguera}
G.~Reguera, K.~D. McCarthy, T.~Mehta, J.~S. Nicoll, M.~T. Tuominen, and D.~R.
  Lovley, ``{Extracellular electron transfer via microbial nanowires},''
  \emph{Nature}, vol. 435(7045), pp. 1098--1101, 2005.

\bibitem{Rustom}
A.~Rustom, R.~Saffrich, I.~Markovic, P.~Walther, and H.-H. Gerdes,
  ``{Nanotubular Highways for Intercellular Organelle Transport},''
  \emph{Science}, pp. 1007--1010, Feb. 2004.

\bibitem{Remis}
\BIBentryALTinterwordspacing
J.~P. Remis, D.~Wei, A.~Gorur, M.~Zemla, J.~Haraga, S.~Allen, H.~E. Witkowska,
  J.~W. Costerton, J.~E. Berleman, and M.~Auer, ``{Bacterial social networks:
  structure and composition of Myxococcus xanthus outer membrane vesicle
  chains},'' \emph{Environmental Microbiology}, vol.~16, no.~2, pp. 598--610,
  2014. [Online]. Available: \url{http://dx.doi.org/10.1111/1462-2920.12187}
\BIBentrySTDinterwordspacing

\bibitem{Kuran5962989}
M.~Kuran, H.~Yilmaz, T.~Tugcu, and I.~Akyildiz, ``{Modulation Techniques for
  Communication via Diffusion in Nanonetworks},'' in \emph{IEEE International
  Conference on Communications (ICC)}, June 2011, pp. 1--5.

\bibitem{Lane}
N.~Lane, ``{Why Are Cells Powered by Proton Gradients?}'' \emph{Nature
  Education}, vol. 3(9):18, 2010.

\bibitem{smith2003foundations}
W.~Smith and J.~Hashemi, \emph{Foundations of Materials Science and
  Engineering}, ser. McGraw-Hill series in materials science and
  engineering.\hskip 1em plus 0.5em minus 0.4em\relax McGraw-Hill, 2003.

\bibitem{Liu2}
H.~Liu, G.~Newton, R.~Nakamura, K.~Hashimoto, and S.~Nakanishi,
  ``{Electrochemical Characterization of a Single Electricity-Producing
  Bacterial Cell of Shewanella by Using Optical Tweezers},'' \emph{Angewandte
  Chemie International Edition}, vol.~49, pp. 6596--6599, Sep. 2010.

\bibitem{Ozalp}
V.~Ozalp, P.~T.R., N.~L.J., and O.~L.F., ``{Time-resolved measurements of
  intracellular ATP in the yeast Saccharomyces cerevisiae using a new type of
  nanobiosensor},'' \emph{Journal of Biological Chemistry}, pp.
  37\,579--37\,588, Nov. 2010.

\bibitem{Golub}
G.~H. Golub and C.~F. Van~Loan, \emph{{Matrix computations (3rd ed.)}}.\hskip
  1em plus 0.5em minus 0.4em\relax Baltimore, MD, USA: Johns Hopkins University
  Press, 1996.

\bibitem{Boyd}
S.~Boyd and L.~Vandenberghe, \emph{{Convex Optimization}}.\hskip 1em plus 0.5em
  minus 0.4em\relax New York, NY, USA: Cambridge University Press, 2004.

\bibitem{Boggs}
P.~Boggs, P.~Domich, and J.~Rogers, ``\BIBforeignlanguage{English}{{An interior
  point method for general large-scale quadratic programming problems}},''
  \emph{\BIBforeignlanguage{English}{Annals of Operations Research}}, vol.~62,
  no.~1, pp. 419--437, 1996.

\bibitem{Biochemistry}
J.~M. Berg, J.~L. Tymoczko, and L.~Stryer, \emph{{Biochemistry}}, 5th~ed.\hskip
  1em plus 0.5em minus 0.4em\relax W H Freeman, 2002.

\bibitem{Teusink}
B.~Teusink, J.~Diderich, H.~Westerhoff, K.~Van~Dam, and M.~Walsh,
  ``{Intracellular glucose concentration in derepressed yeast cells consuming
  glucose is high enough to reduce the glucose transport rate by 50\%},''
  \emph{Journal of Bacteriology}, vol. 180(3), pp. 556--562, Feb. 1998.

\bibitem{Wang}
P.~Wang, L.~Robert, J.~Pelletier, W.~Dang, F.~Taddei, A.~Wright, and
  S.~Junemail, ``{Robust Growth of \emph{Escherichia coli}},'' \emph{Current
  Biology}, vol.~20, pp. 1099--1103, June 2010.

\end{thebibliography}

\begin{IEEEbiography}[{\includegraphics[width=1in,height=1.25in,clip,keepaspectratio]{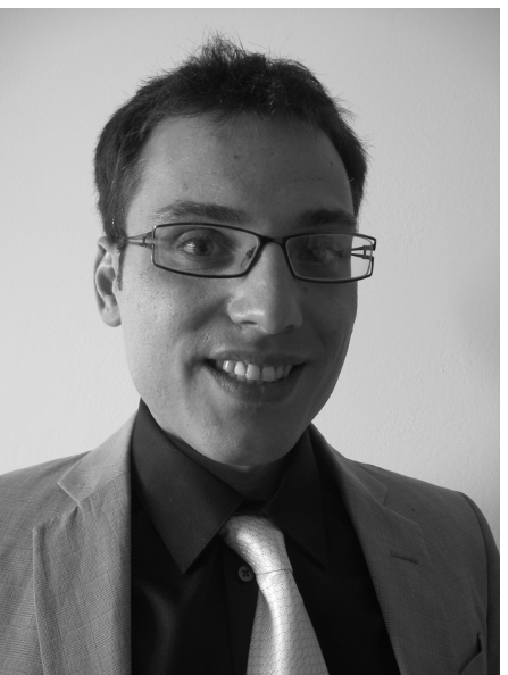}}]
{Nicol\`{o}~Michelusi} (S'09, M'13) received the B.Sc. (with honors), M.Sc.
 (with honors) and Ph.D. degrees from the University of Padova, Italy, in 2006, 2009 and 2013, respectively,
and the M.Sc. degree in Telecommunications Engineering from the Technical University of Denmark in 2009, as part of the T.I.M.E. double degree program.
In 2011, he was at the University
of Southern California, Los Angeles, USA,
and, in Fall 2012, at Aalborg University, Denmark, as a visiting research scholar.
He is currently a post-doctoral research fellow at the Ming Hsieh Department of Electrical Engineering, University of Southern California, USA.
 His research interests lie in the areas of
wireless networks, stochastic optimization, distributed estimation and modeling of bacterial networks.
  Dr. Michelusi  serves as a reviewer for the IEEE Transactions on Communications, IEEE Transactions on Wireless Communications, IEEE Transactions on Information Theory, IEEE Transactions on Signal Processing, IEEE Journal on Selected Areas in Communications,
 and IEEE/ACM Transactions on Networking.
 \end{IEEEbiography}

 \begin{IEEEbiography}[{\includegraphics[width=1in,height=1.25in,clip,keepaspectratio]{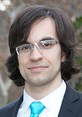}}]
{Sahand Pirbadian} is a Ph.D. candidate in the Department of Physics and Astronomy at the University of Southern California. He received his B.Sc. degree in physics from Sharif University of Technology, Iran in 2010. His research interests include extracellular electron transfer (EET), electron transport in bacterial communities, bacterial multiheme cytochromes and bacterial nanowires.
 \end{IEEEbiography}

 \begin{IEEEbiography}[{\includegraphics[width=1in,height=1.25in,clip,keepaspectratio]{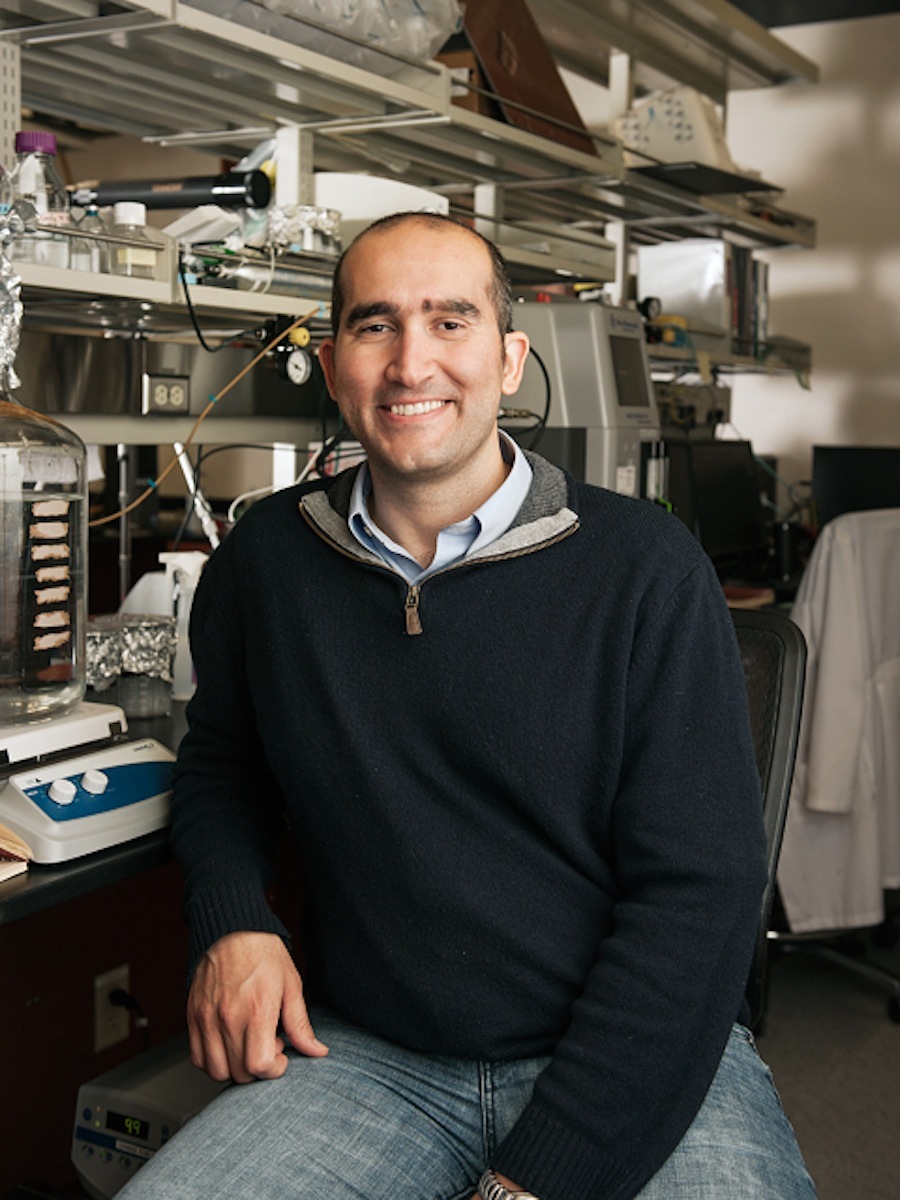}}]
{Moh El-Naggar}  is an assistant professor of physics at the University of Southern California's Dornsife College of Letters, Arts and Sciences. El-Naggar studies energy conversion and charge transmission at the interface between living cells and natural or synthetic surfaces. In 2012, he was named one of Popular ScienceÕs ÔBrilliant 10Õ, and in 2014 he was awarded the Presidential Early Career Award for Scientists and Engineers (PECASE). El-NaggarÕs work has important implications for cell physiology and astrobiology: it may lead to the development of new hybrid materials and renewable energy technologies that combine the exquisite biochemical control of nature with the synthetic building blocks of nanotechnology. El-Naggar earned his B.S. degree in mechanical engineering from Lehigh University (2001), and his Ph.D. in engineering and applied science from the California Institute of Technology (2007).
 \end{IEEEbiography}
 
\begin{IEEEbiographynophoto}{Urbashi Mitra}
  received the B.S. and the M.S. degrees from the University of California at Berkeley and her Ph.D. from Princeton University. After a six year stint at the Ohio State University, she joined the Department of Electrical Engineering at the University of Southern California, Los Angeles, where she is currently a Professor. She is a member of the IEEE Information Theory Society's Board of Governors (2002-2007, 2012-2014) and the IEEE Signal Processing SocietyÕs Technical Committee on Signal Processing for Communications and Networks (2012-2014). Dr. Mitra is a Fellow of the IEEE.  She is the recipient of:  2012 Globecom Signal Processing for Communications Symposium Best Paper Award, 2012 NAE Lillian Gilbreth Lectureship, USC Center for Excellence in Research Fellowship (2010-2013), the 2009 DCOSS Applications \& Systems Best Paper Award, Texas Instruments Visiting Professor (Fall 2002, Rice University), 2001 Okawa Foundation Award, 2000 OSU College of Engineering Lumley Award for Research, 1997 OSU College of Engineering MacQuigg Award for Teaching, and a 1996 National Science Foundation (NSF) CAREER Award. Dr. Mitra currently serves on the IEEE Fourier Award for Signal Processing committee and the IEEE James H. Mulligan, Jr. Education Medal committee. She has been/is an Associate Editor for the following IEEE publications: Transactions on Signal Processing (2012--), Transactions on Information Theory (2007-2011), Journal of Oceanic Engineering (2006-2011), and Transactions on Communications (1996-2001). She has co-chaired: (technical program) 2014 IEEE International Symposium on Information Theory in Honolulu, HI,  2014 IEEE Information Theory Workshop in Hobart, Tasmania, IEEE 2012 International Conference on Signal Processing and Communications, Bangalore India, and  the IEEE  Communication Theory Symposium at ICC 2003 in Anchorage, AK;  and  general co-chair for the first ACM Workshop on Underwater Networks at Mobicom 2006, Los Angeles, CA Dr. Mitra was the Tutorials Chair for IEEE ISIT 2007 in Nice, France and the Finance Chair for IEEE ICASSP 2008 in Las Vegas, NV. Dr. Mitra has held visiting appointments at: the Delft University of Technology, Stanford University, Rice University, and the Eurecom Institute. She served as co-Director of the Communication Sciences Institute at the University of Southern California from 2004-2007.  Her research interests are in: wireless communications, communication and sensor networks, detection and estimation and the interface of communication, sensing and control.
  \end{IEEEbiographynophoto}

\end{document}